# Elasticity Theory and Shape Transitions of Viral Shells.

Nguyen Toan[1,2], Robijn F. Bruinsma[1], and William M. Gelbart[2]

Department of Physics and Astronomy[1], and Department of Chemistry and Biochemistry[2],

The University of California at Los Angeles, Los Angeles 90049, California

**ABSTRACT**

**Recently, continuum elasticity theory has been applied to explain the shape transition of icosahedral viral capsids – single-protein-thick crystalline shells – from spherical to "buckled"/faceted as their radius increases through a critical value determined by the competition between stretching and bending energies of a closed 2D elastic network. In the present work we generalize this approach to capsids with non-icosahedral symmetries, e.g., spherocylindrical and conical shells. One key new physical ingredient is the role played by nonzero spontaneous curvature. Another is associated with the special way in which the energy of the twelve topologically-required five-fold sites depends on the "background" local curvature of the shell in which they are embedded. Systematic evaluation of these contributions leads to a shape "phase" diagram in which transitions are observed from icosahedral to spherocylindrical capsids as a function of the ratio of stretching to bending energies and of the spontaneous curvature of the 2D protein network. We find that the transition from icosahedral to spherocylindrical symmetry is continuous or weakly first-order near the onset of buckling, leading to extensive *shape degeneracy*. These results are discussed in the context of experimentally observed variations in the shapes of a variety of viral capsids.**

# I) Introduction

The genetic information of a virus is surrounded by a closed shell of protein molecules, the *capsid*, which protects the enclosed RNA or DNA genome molecules against enzymatic digestion[1]. Capsids are also exceptionally resilient under applied mechanical forces. At the same time, a capsid must direct the efficient release of the genome molecules into prospective host cells. It is not surprising that the synthesis of artificial protein cages that can reproduce such remarkable properties is a rapidly developing area of materials science, and yet the relevant design criteria are only beginning to be understood[2,3].

Most capsids have either a sphere-like or a rod-like morphology. Modern methods of X-ray crystallography and Cryo-TEM tomography allow the reconstruction of the sphere-like viral shells with near-atomic resolution[4]. Sphere-like shells have, nearly always, the symmetry of an *icosahedron*. In many cases, the proteins (or "subunits") that constitute the shell can be grouped into "capsomers", e.g., oligomers constructed from either five ("pentamer") or six ("hexamer") subunits. Pentamers are located on twelve equidistant sites that form the vertices of an icosahedron. The number of hexamers that constitute the faces of the icosahedron adopt certain "magic" numbers given by 10 (T-1), with T an integer index equal to 1, 3, 4, 7,.... Remarkably, there are many instances in which these intricately patterned icosahedral viral shells assemble *spontaneously* under appropriate *in vitro* conditions[5].

Over forty years ago, Caspar and Klug (CK) showed in a seminal paper how the "T Number" sequence of structures could be obtained from simple geometric considerations[6]. They constructed equilateral triangles with vertices located at the centers of a two-dimensional hexagonal lattice (see Fig.1B). If a triangle has one of its vertices at the origin then it can be indexed by the pair of integers h and k that determine the location of one of the two other vertices in terms of the two basis vectors of the hexagonal lattice (Fig.1B). The icosahedron is then constructed from a folding template of twenty of such triangles, replacing a hexagon by a pentagon at each of the twelve vertices of the icosahedron (see Figs.1A and 1C). The CK construction has ever since remained the structural basis for the classification of "spherical" viral capsids.

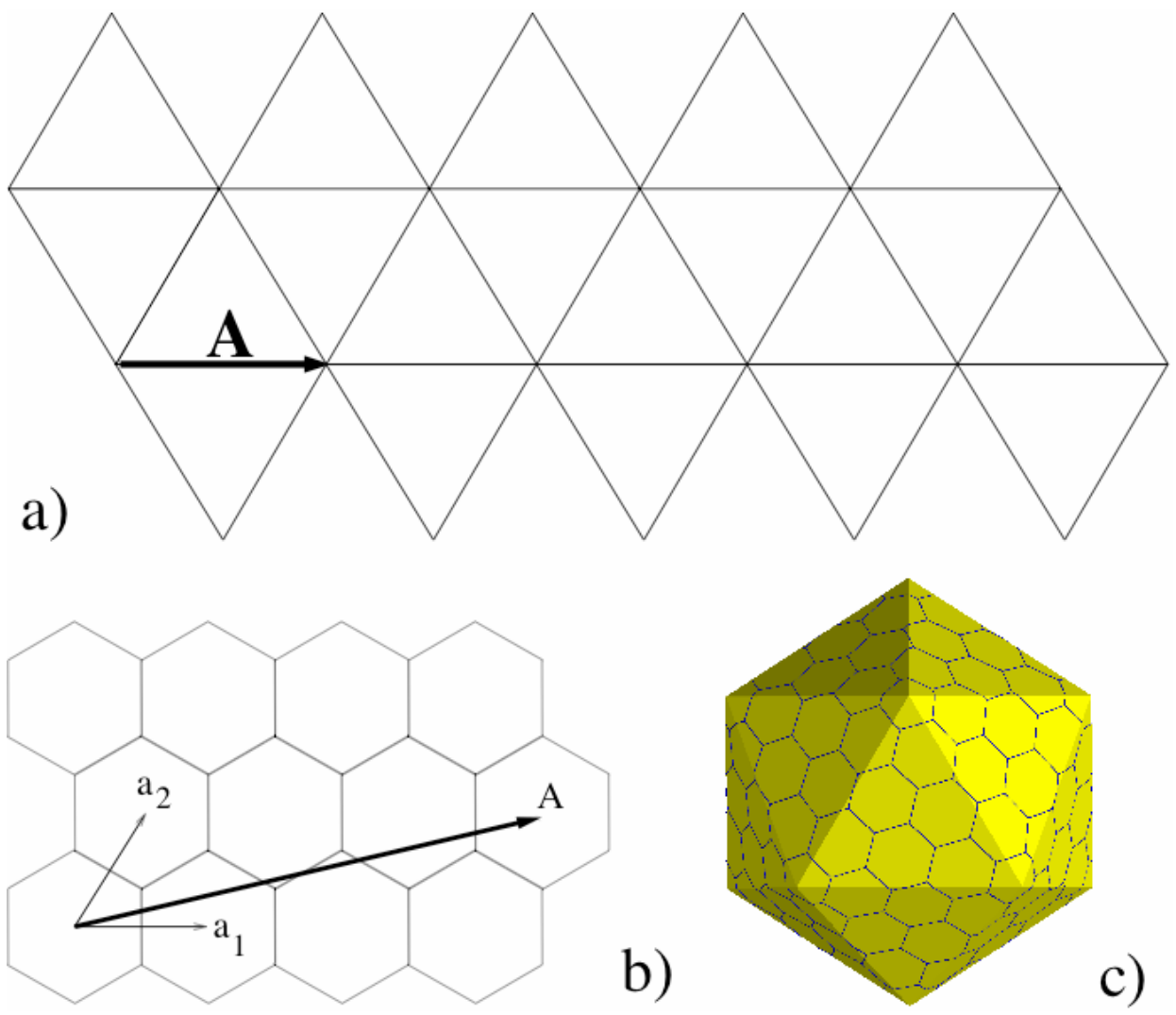

Figure 1: CK construction of icosahedral shells. Fig.1A: Folding template for an icosahedron consisting of twenty equilateral triangles. The triangles are indexed by a lattice vector $\vec{A} = h\vec{a}_1 + k\vec{a}_2$ of a hexagonal lattice with basis vectors $\hat{a}_1$ and $\hat{a}_2$. Fig.1B shows the case h=3 and k=1. Fig.1C shows an icosahedron obtained from folding the template for this lattice vector, which corresponds to $T=h^2+k^2+hk=13$. Note that there are six hexagons for each face of the icosahedron**,** and that there are 10(T-1)=120 hexagons in total.

The CK icosahedra are isometric in the strict sense that the construction does not change the distance between two sites of the original hexagonal lattice[7]. Though an icosahedral shell constructed from an inextensible hexagonal sheet indeed must be isometric, actual protein materials do support elastic strain. If an icosahedral shell is constructed from a hexagonal sheet that does support elastic strain, then the bending energy cost of the sharp edges of the CK icosahedron (see Fig.1C) can be relieved by allowing the triangular faces of the icosahedron to bulge out. The resulting elastic stretching will be referred to as the "in-plane" elastic stress that must be balanced with the "out-of-plane" bending energy of the sheet along the edges.

The theoretical prediction of the structure of a viral capsid is, in general, a daunting problem in view of the complex internal structure of the subunits. However, for capsids with a very large numbers of subunits, one expects that the capsid can be described by the *continuum theory of elasticity*. In continuum elasticity theory, the physical properties of a shell are determined by only a few phenomenological constants such as the 2D Young's Modulus Y of the sheet and the Helfrich bending constant κ. (The actual values of these constants of course still depend on molecular-level interactions between subunits.) Lidmar, Mirny, and Nelson[8] (LMN) have developed such a continuum description for icosahedral capsids and determined shell shapes that minimize the sum of the bending and stretching energy costs. In this approach the twelve pentagons of the CK construction are replaced by twelve 5-fold *disclination defects*, each of which is surrounded by a field of elastic stress. The energy of an icosahedral shell of area S depends only on the single dimensionless quantity $\gamma = \frac{YS}{\kappa}$, the ratio of stretching and bending energies, known as the *Föppl-von Kármán (FvK) Number*.[9] Figure 2 shows the continuum theory elastic energy of an icosahedral shell – obtained by the numerical energy minimization described in Section IV – as a function of the FvK Number.

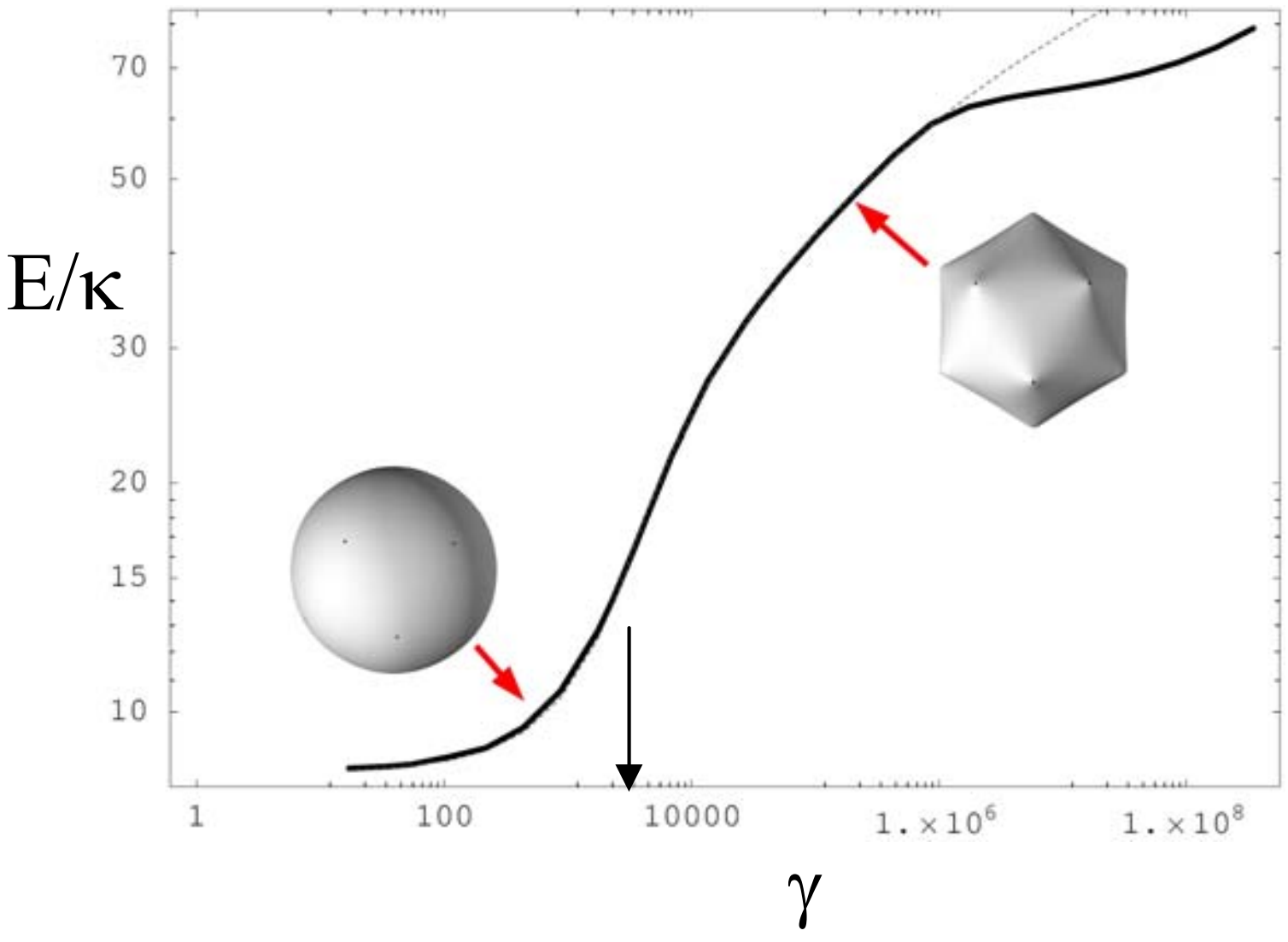

Figure 2: Elastic energy E of an icosahedral shell expressed in units of the bending constant κ, for the case of zero spontaneous curvature. The horizontal axis is the Föppl-von Kármán Number $\gamma = \frac{YS}{\kappa}$ with S the surface area of the shell, Y the Young's

Modulus, and $\kappa$ the bending constant. For FvK numbers near a critical value $\gamma_B$ around 3000 (arrow) a *buckling transition* takes place, with the shell shape transforming from spherical to icosahedral. The dotted line shows the result of a fit to the LMN theory (see Section 3C). On a linear scale, E(S) has a negative curvature for FvK numbers significantly above the buckling threshold.

The shell shape changes from (nearly) spherical to (nearly) icosahedral at a *buckling transition* for $\gamma$ values near a critical value $\gamma_B$ around 3000. LMN showed that the theory can account rather well for capsid structures of certain of the larger viruses. More generally, they noted that larger viruses are noticeably more polyhedral than smaller viruses, in accordance with the theory. An actual buckling transition has been reported to take place during the expansion/maturation of the T=7 HK97 viral capsid[10,11].

The primary aim of the present paper is the extension of the continuum theory of viral shells to include *non-icosahedral* viruses and to construct a shape phase-diagram. Our focus is centered on non-icosahedral viruses with capsid structures based on a hexamer-pentamer organization that is obtainable from a generalization of the CK construction. Shells that belong to this class – which specifically excludes the open-ended cylindrical viruses like TMV – are *spherocylindrical* ones commonly found among the bacteriophage viruses, such as certain "T-even" phages (plus their mutants), as well as the $\phi$CbK and $\phi$29 bacteriophages. The capsids of these viruses consist of two half-icosahedral caps connected by an elongated, cylindrical, mid-portion composed of a ring of hexamers[12]. This "buckytube" structure is also encountered as a variant of the T = 7 Papovaviruses[13]. Similarly, point mutations in capsid proteins may transform an icosahedral T-Number shell into a tubular shell of variable length[14]. Interesting in this context are the *polymorphic* viruses, i.e., viruses whose capsids can exhibit *both* spherical and tubular morphologies. Self-assembly studies of solutions of the capsid proteins of the Cowpea Chlorotic Mottle Virus (CCMV)[15] and the Polyoma/SV40 animal virus[16] – without, respectively, their RNA or DNA genome molecules – report both sphere-like and tubular structures with, for the CCMV case, the relative abundance dependent on the pH level and salt concentrations. The Alfalfa Mosaic Virus (AMV) is *naturally* polymorphic, with its multipartite genome – RNA molecules of different lengths – separately encapsidated by extended shells of various lengths while self-assembly without the genome molecules will produce T=1 icosahedral shells[17]. Finally, the Human Immunodeficiency Virus (HIV) shows still broader polymorphism in its capsid shape, including cone-like structures in addition to tubes and roughly spherical ones[18].

Our description will be based on the LMN continuum theory of elastic shells, but generalized to non-spherical shapes and including the concept of *spontaneous curvature*, already proposed by CK as a central determinant for capsid assembly. This generalization was motivated by a detailed structural study of CCMV capsids[19] that suggested the competition between tubular and spherical geometries might be controlled by two biophysical effects.

The first effect concerns the *asymmetry* of viral subunits and capsomers with respect to the interior and the exterior of a capsid. One aspect of this asymmetry is the fact that CCMV capsid subunits are joined with a preferred nonzero angle, along two-fold contacts, which maximizes the number of hydrophobic side-groups that are shielded from the surrounding aqueous environment. Next, charged residues facing the viral exterior are usually negatively charged while those facing the interior are mostly positively charged. The result of this "in-out" asymmetry is that a hexagonal sheet of CCMV capsid proteins sheet in general has a certain preferred curvature determined by the ambient conditions. For the case of CCMV, this preferred – spontaneous – curvature is strongly dependent on the concentration of divalent ions. Size control by spontaneous curvature in CCMV and other T=3 RNA viruses is associated with *conformational switching*[20], since subunits that participate in two-fold contacts must adopt different conformations depending on whether the contact is flat or bent. This conformational switch can be a terminal protein segment that is either ordered or disordered[21]. Upon removal of this switch, the capsid proteins form minimal sized T=1 shells. Similar conformational switching has been observed for a T=7 virus[22]. Whether size control of *large* viruses can also proceed through spontaneous curvature, involving both spherical and non-spherical shapes, will be one of the important issues of this paper.

The second effect noted in the CCMV study is related to the *energy difference* between pentamers and hexamers. If the energy cost of a pentamer is comparable to that of a hexamer, then the minimum energy structure would be expected to be an icosahedral shell with a radius of order the inverse of the preferred curvature. If however the energy cost of a pentamer is large compared to that of a hexamer, then a single long tubular structure with a radius of order the inverse curvature should have a lower energy than a group of icosahedral shells with the same total number of subunits, because the tube has a lower ratio of pentamers over hexamers. We note, in this context, that similar arguments[23] have been shown to account for the preference of rod versus sphere shapes of surfactant micellar aggregates, with "cap" and "body" packing taking the place of pentamers and hexamers, respectively.

In continuum elasticity theory, the pentamer/hexamer energy difference is in fact included in the form of the "core energy" of the disclination defects that is determined by the elastic constants, though it should be noted that the (free) energy difference between pentamer and hexamer oligomers in actual capsids is likely to involve as well a conformational switching energy that is not related to the elastic constants of the shell. However, the first effect – preferential curvature – has not yet been included in any way whatsoever in the continuum theory of shells. We will denote the preferred mean curvature of a shell by $C_0$, so the inverse $1/C_0$ is the spontaneous-curvature radius that should determine the size scale of a capsid in a self-assembly experiment. There are now two characteristic length scales in the problem: the spontaneous-curvature radius $1/C_0$ and the buckling radius $R_B$ defined by the critical value $\gamma_B = \frac{4\pi R_B^2 Y}{\kappa}$ of the ratio of stretching to bending energies. Note by the way that, at least *a priori*, the buckling radius also could act as a size scale for capsids, and we will in fact see that that is a real possibility.

The preferred-curvature concept should play an important role in the spontaneous self-assembly of capsid shells from a solution of subunits (or oligomers of subunits). If we view Fig.2 as a plot of the energy E(S) versus the (2D) system size S – since the FvK Number is proportional to S – then E(S) is seen to have no minima, and a *negative curvature* for all FvK Numbers above $10^4$. For conventional many-body systems, a negative curvature of the free energy as a function of system size would signal some form of phase separation. We will show that negative curvature of E(S) leads to polydispersity of the size distribution in a self-assembly experiment. One might expect that the spontaneous curvature effect could overcome this negative curvature "problem" and produce a reasonably monodisperse distribution of capsid sizes having an area of order $1/C_0^2$. It should be noted in this context that spontaneous curvature is not the only form of size control in viral assembly. Many large viruses employ a *scaffold structure*, i.e., a condensation surface for the subunits that may disassemble afterwards, while in other cases, such as the Polyoma/SV40 virus, the genome itself appears to act as a size gauge.

A more specific aim of the paper involves the application of the continuum theory to the *retroviruses*. The capsid shells of retroviruses are constituted from a rather large numbers of hexamers and pentamers, of the order of 300, and continuum theory is expected to be applicable. Retrovirus capsids usually do not exhibit icosahedral symmetry, but they can be spherical, such as the murine leukemia virus capsid, or tubular such as the Mason-Pfizer monkey virus[24]. Particularly interesting are the capsids of the HIV-1 virus, the majority of which have a *conical* shape, while a smaller fraction has a tubular structure[18,25]. Recent cryoTEM tomography studies[18] confirm that the HIV-1 conical shells are *polydisperse*, i.e., with a variety of sizes and shapes. Conical HIV-1 capsids will form by self-assembly under *in vitro* conditions[26] - in the presence of the viral RNA genome molecules - indicating that this shell structure really may be a minimum of the free energy, though a range of other non-spherical self-assembled structures are encountered as well – apart from cones – such as spheres, spherocylinders, and curved sheets. Whether scaffolding plays a role in natural HIV-1 assembly is currently not known.

Although the generalized continuum theory should provide a description for the self-assembly of icosahedral and spherocylindrical shells, the HIV-1 conical shells do pose a serious challenge. In the literature on *lipid bilayers*[27], a similar continuum theory – including the spontaneous curvature effect but excluding the in-plane elasticity – has been applied with success to describe the shape of closed *fluid* surfaces. The resulting shape catalogue actually does include, apart from spheres and tubes, conical shaped ("pear shaped") shells. However, both the surface area S and the enclosed volume V of lipid vesicles are essentially fixed (the latter by osmotic pressure). Although the surface area of a capsid can be assumed fixed by the number of capsomers – the T Number – the enclosed volume V of a viral shell is *not* a fixed quantity. Capsid shells are permeable to water molecules and to small salt ions, so the osmotic pressure difference between the exterior and interior of an empty viral shell must be zero. Without the fixed-volume constraint, pear-shaped vesicles would not be stable. The reason is that the curvature of a cone changes continuously along the cone axis. If the *spatial average* of the curvature of the cone is set equal to the preferred curvature $C_0$ of the proteins, then a cylinder still would have a lower bending energy since the curvature of a cylinder is constant and can be equated everywhere to $C_0$. If this argument would hold true as well for shells that support elastic stress, then continuum theory would not allow for the spontaneous self-assembly of empty, conical HIV-1 shells.

We will show that the shape of elastic shells with preferred curvature is determined by the competition of the spontaneous curvature effect with two different physical mechanisms. The first is the effective *pair potential* between the 5-fold sites. This pair potential is logarithmic over a large range of FvK Numbers and is responsible for the negative curvature of the E(S) plot of Fig.2. The negative curvature of the pair potential has a tendency to drive a *decomposition* of the twelve 5-fold sites, thereby favoring non-spherical shell shapes. The second mechanism, which will be termed the "*misfit energy*", is the deformation energy cost of imbedding a 5-fold disclination site into the curved "background" surface of the various competing structures.

The key physical results that result from this competition are as follows:

i). If the dimensionless spontaneous curvature $\alpha = C_0 S^{1/2}$ is small compared to unity, then the icosahedral shell structure has, for *any* FvK Number $\gamma$, a lower elastic energy than that of either the cone or the spherocylinder. For large FvK Numbers, stability of the icosahedral shell is provided by its low misfit energy, which overcomes the "negative-curvature" effect mentioned above.

ii). Self-assembly of icosahedral shells with a well-defined size determined by the spontaneous curvature radius requires the FvK Number $\gamma$ to be below the buckling threshold $\gamma_B$. Self-assembly of monodisperse shells above the buckling threshold evidently requires a scaffold-type mechanism.

iii). For FvK Numbers near or below the buckling threshold, and for spontaneous curvatures $C_0$ of order 1/R, *there is a substantial portion of the phase diagram where the energies of the sphere, cone, and tube approach each other to within 0.5% of the total* *elastic energy.* We will argue that this energy difference is of the order of the thermal energy $k_BT$ and present evidence from the literature on in vitro self-assembly for the existence of extensive shape diversity in the transition region between sphere and spherocylinder. The observations on HIV-1 self-assembly would indicate that its capsid formation is naturally located in precisely this section of the phase-diagram. We will, however, discuss certain difficulties with this interpretation in the Discussion.

iv). For FvK numbers that are significantly above the buckling threshold, and for $C_0R$ somewhat larger than unity, there is a weakly "first-order" transition from an icosahedral to a tubular shell, while for FvK numbers well below the transition is strongly first-order.

The paper is organized as follows. In Section II, we generalize the CK construction to allow for a unified isometric description of icosahedral, spherocylindrical and conical structures. In Section III, we present a simple analytical description of non-icosahedral shells in which we feature the dependence of disclination energies on the curvature of the hexagonal lattice in which they are embedded. In Section IV we use numerical energy minimization to determine the "shape" phase diagram shown in Fig.13, at the heart of which is a region of intermediate capsid size and spontaneous curvature where a number of different shapes, including conical, are found to have comparable energies. The implications of our results for capsid assembly in general and HIV-1 in particular are discussed in the concluding Section V.

# II) Non-Icosahedral Isometric Shells

Here we generalize the CK construction in order to determine the location of the twelve 5-fold sites for non-icosahedral isometric shells. These isometric structures will form a starting platform for the analytical and numerical studies of Sections III and IV.

The folding template for the classical CK construction of icosahedral shells was already shown in Fig. 1A. The vector $\vec{A}$ determining the folding template is a hexagonal lattice vector:

$$\vec{A} = h\vec{a}_1 + k\vec{a}_2, \tag{2.1}$$

with h and k a pair of non-negative integers and $\vec{a}_1$ and $\vec{a}_2$ basis vectors of the hexagonal lattice (see Fig.1B). A hexameric capsomer is associated with each lattice site. After folding the template into an icosahedron (see Fig.1C), the final number N(h,k) of hexamers and pentamers equals:

$$N(h,k) = 10\left(h^2 + k^2 + hk\right) + 2, \tag{2.2}$$

which is usually written as N(h, k) = 10 T (h,k) + 2 with $T(h,k) = h^2 + k^2 + hk$.

The construction of an isometric spherocylinder starts in the same fashion. The folding template for a spherocylinder is shown in Fig.3A.

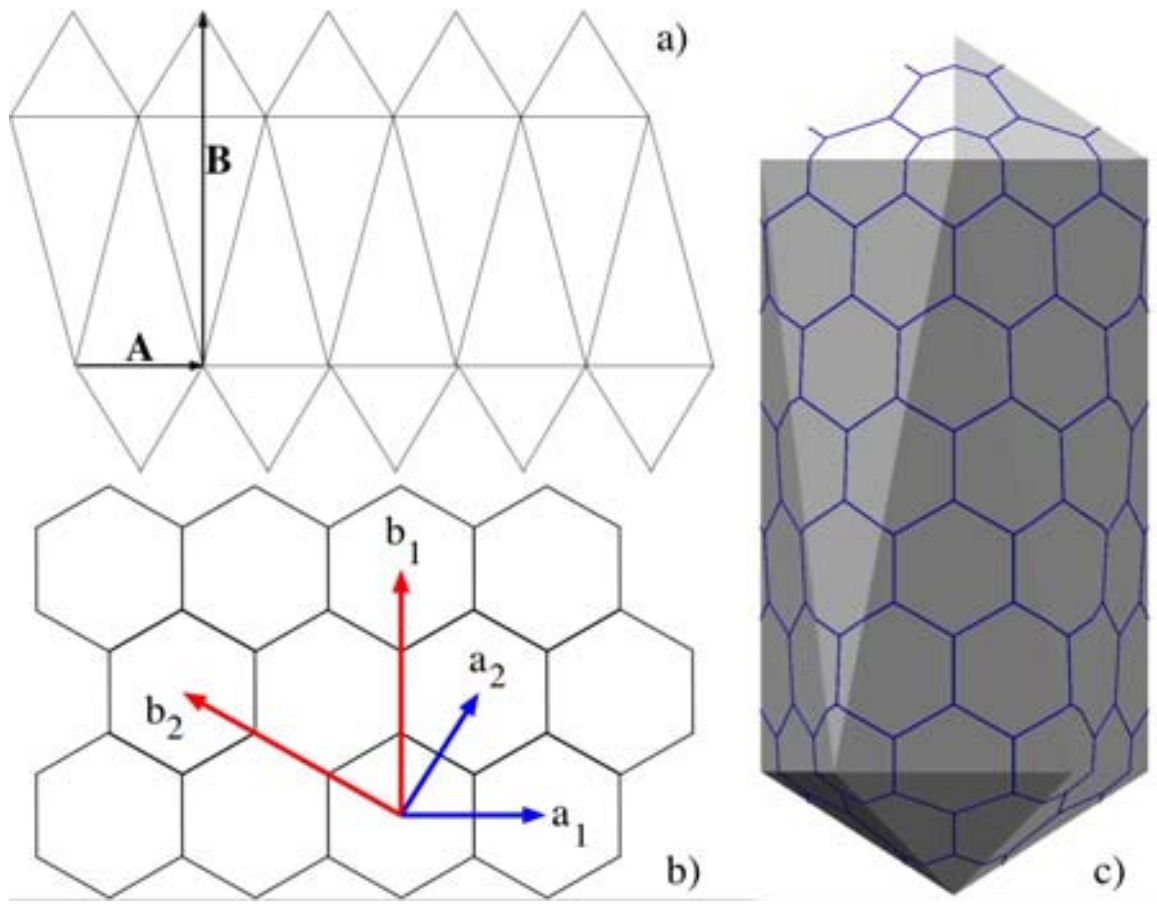

Figure 3: Construction of an Isometric Spherocylinder. Fig.3A: Folding template for an isometric spherocylinder. Fig. 3B: The basis vectors of the template are $\vec{A} = n(h\vec{a}_1 + k\vec{a}_2)$ and $\vec{B} = m(h\vec{b}_1 + k\vec{b}_2)$. They are perpendicular to another with (h,k) and (n,m) any two pairs of non-negative integers with m > n. For m=n, the spherocylinder reduces to an icosahedron. Fig.3C: Isometric spherocylinder with n=2,m=0 and h=3,k=0.

First, we define a lattice vector $\vec{A} = h\vec{a}_1 + k\vec{a}_2$ for the folding of the two (semi) icosahedral capping sections of the template. In the simplest case, the two caps are displaced along a direction perpendicular to $\vec{A}$ by the lattice vector $\vec{B}$, which is defined by the location of the pentamer site of the tip of one of the caps of the spherocylinder starting from one of the pentamers on the base of the opposite cap (see Fig.3A). All lines shown in the folding template of Fig.3 must be lattice vectors as well. If we set $\vec{B} = p\vec{a}_1 + q\vec{a}_2$, then $\vec{A}$ is perpendicular to $\vec{B}$ if $p/q = -(h+2k)/(2h+k)$. This condition is satisfied by choosing $\vec{B} = h\vec{b}_1 + k\vec{b}_2$, with

$$\vec{b}_1 = -\vec{a}_1 + 2\vec{a}_2, \qquad (2.3a)$$

$$\vec{b}_2 = -2\vec{a}_1 + \vec{a}_2. \qquad (2.3b)$$

As shown in Fig.3B, the $\vec{b}_{1,2}$ vectors are perpendicular to the $\vec{a}_{1,2}$ lattice basis vectors. In fact, if we multiply $\vec{A} = h\vec{a}_1 + k\vec{a}_2$ by an arbitrary integer n and $\vec{B}$ by an integer m, then $\vec{B}$ remains perpendicular to $\vec{A}$. The folding template of a spherocylinder is thus defined by two pairs of basis vectors [$(\vec{a}_1, \vec{a}_2)$ and $(\vec{b}_1, \vec{b}_2)$] and two pairs of integer [(h,k) and (m,n)]:

$$\vec{A} = n(h\vec{a}_1 + k\vec{a}_2), \qquad (2.4a)$$

$$\vec{B} = m(h\vec{b}_1 + k\vec{b}_2). \qquad (2.4b)$$

The total number of capsomers of a spherocylinder defined by the two pairs (h,k) and (n,m) is:

$$N(h,k\,|\,n,m) = 10mn(h^2 + k^2 + hk) + 2\,. \tag{2.5}$$

For the case of the icosahedron, with m = n, this reduces to Eq.2.2. This construction is not the most general case, since we could have chosen the lattice vector $\vec{B}$ to lie along a direction that is not perpendicular to $\vec{A}$, which would have produced a *helical* spherocylinder.

In order to construct an isometric cone, we start from the isometric spherocylinder and move one or more pentamers from the top end cap to the bottom end cap. We will restrict ourselves to the "5-7" case with the smaller cap containing five pentamers and the larger cap containing seven pentamers. The corresponding folding template is shown in Fig.4A.

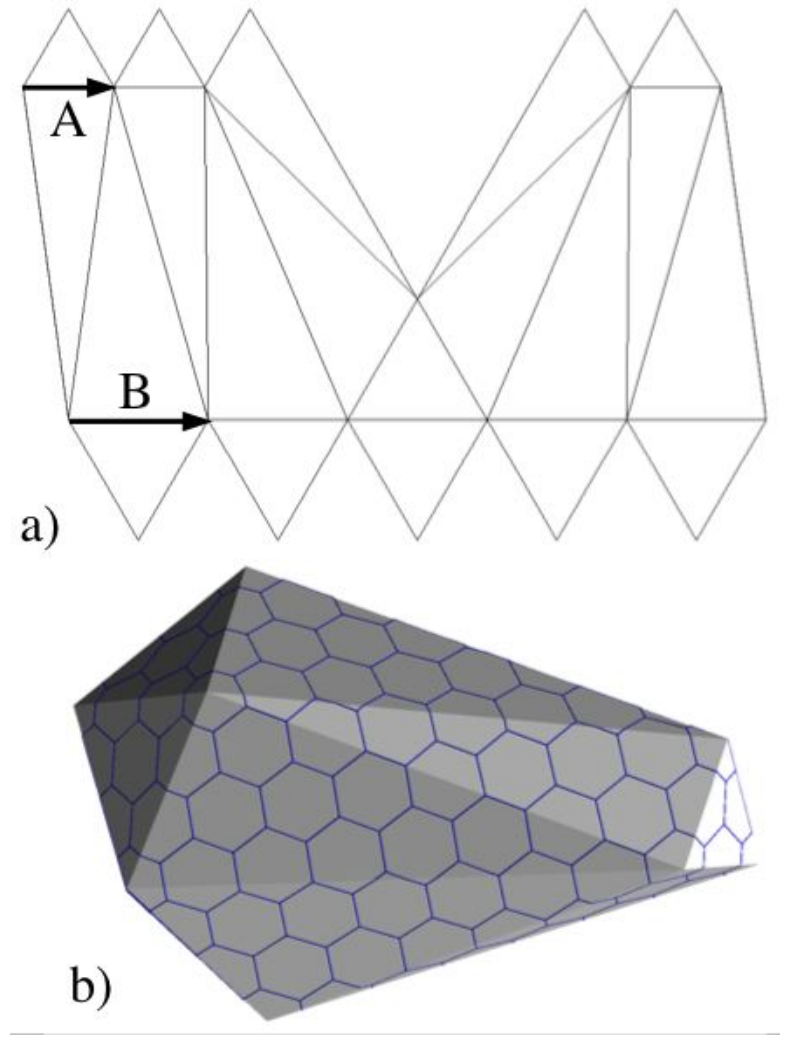

Figure 4: Construction of an Isometric Cone. Fig.4A: Folding template of an Isometric Cone. The two lattice vectors $\vec{A} = n(h\vec{a}_1 + k\vec{a}_2)$ and $\vec{B} = m(h\vec{a}_1 + k\vec{a}_2)$ are parallel, with (h,k) and (m,n) any two pairs of non-negative integers with m > n. Fig.4B: Isometric Cone with h=1, k=0, m=3, and n=2.

The top and bottom caps can be considered as two sections of an isometric icosahedron with T numbers $T^S$ and $T^1$ and folding lattice vectors $\vec{A}$ and $\vec{B}$. All directions shown in the folding template must again be lattice vectors. Let the T number of the top cap be

T(mh, mk) and that of the bottom cap T(nh, nk). The corresponding lattice vectors are, respectively:

$$\begin{aligned} \vec{A} &= n(h\vec{a}_1 + k\vec{a}_2) \\ \vec{B} &= m(h\vec{a}_1 + k\vec{a}_2). \end{aligned} \tag{2.6}$$

The size ratio of the two caps is the size ratio of the lengths of these two vectors, i.e., m/n, which can be any rational number greater than or equal to one. Varying this size ratio corresponds to varying the distance between the top and bottom caps. The total number of capsomers equals:

$$N(h,k \mid m,n) = 10(2m^2 - n^2)(h^2 + k^2 + hk) + 2\ . \tag{2.7}$$

# III) Elasticity Theory of Capsids

The theory of elasticity assigns to thin elastic shells an energy H that can be written as the sum of an "in-plane" stretching energy $H_S$ and an "out-of-plane" bending energy $H_B$. The stretching energy of an elastic sheet of hexagonal symmetry is given by

$$H_S = \frac{1}{2}\int dS\left(2\mu u_{ij}^2 + \lambda u_{ii}^2\right). \tag{3.1}$$

Here, $u_{ij}$ is the strain tensor for displacement within the plane of the shell while $\lambda$ □□□ $\mu$ are two phenomenological constants, known as the "Lamé Coefficients", that are related to the Area Modulus by $B = \lambda + 2\mu$ and to the (2D) Young's Modulus by $Y = \frac{4\mu(\mu+\lambda)}{2\mu+\lambda}$. The hexagonal sheet is assumed to be closed, which by Euler's Theorem requires the introduction of twelve sites having five-fold symmetry, the disclination defects. Recall that, within the continuum theory, disclination defects play the role of the twelve pentamers of the CK construction discussed in Section II.

The out-of-plane bending energy of an elastic shell is given by:

$$H_B = \frac{1}{2}\int dS\left(\kappa\left(H - C_0\right)^2 + 2\kappa_G K\right). \tag{3.2}$$

Here $\kappa$ is the Helfrich bending constant, $H = 1/R_1 + 1/R_2$ is the mean curvature, with $R_1$ and $R_2$ the principal radii of curvature, and $C_0$ is the preferred or spontaneous curvature. In the second term, $K = \frac{1}{R_1 R_2}$ is the Gaussian curvature with $\kappa_G$ the Gaussian bending constant. Within the generalized continuum elasticity theory, a viral shell is thus characterized by five phenomenological constants: the two Lamé Coefficients ($\lambda$ and $\mu$), the two bending moduli ($\kappa$ and $\kappa_G$), and the preferred curvature ($C_0$). We will not assume any *a priori* restrictions on these phenomenological constants until the concluding section, where we will discuss typical ranges as obtained from biophysical and numerical studies.

The minimization of the elastic energy given by Eqs. 3.1 and 3.2 leads to a set of coupled non-linear equations, derived by Föppl and von Kármán, whose solution in general requires numerical methods. In the remainder of this section we will restrict ourselves to certain limiting cases where it is possible to apply analytical methods.

**A) The Helfrich Limit**

In the limit of $\mu = 0$ (and hence of vanishing Y and of $\gamma = YS^2/\kappa$) the elastic energy of the 5-fold disclinations plays no role. We will assume that the Lame Coefficient $\lambda$ is infinite (and hence the Area Modulus B) so the surface area is fixed. The bending energy $H_B$, Eq.3.2, then has to be minimized for fixed total area S. In this limiting case the capsid surface is effectively fluid and, as noted in the Introduction, we can use in this regime the results of the Helfrich theory of lipid bilayers[26] provided we do not maintain the volume as a fixed quantity.

The bending energy $E_s(R)$ for a sphere of radius R, in units of the bending modulus $\kappa$, is – according to Eq. 3.2 – a quadratic function of the preferred curvature $C_0$:

$$E_s(R)/\kappa = D(0) - 8\pi C_0 R + 2\pi C_0^2 R^2 \,, \tag{3.3}$$

with $D(0) = 4\pi\left(2 + \kappa_G/\kappa\right)$. Note that the bending energy is minimized by $E_s/\kappa = 4\pi\kappa_G/\kappa$ when the mean curvature 2/R equals $C_0$. The bending energy of a spherocylinder with radius $\rho$ and height h – see Fig.7B – is given by

$$E_{sc}(\rho,h)/\kappa = E_s(\rho)/\kappa + \pi\left(\frac{h}{\rho}\right) - 2\pi C_0 h + \pi h\rho C_0^2 \,. \tag{3.4}$$

This energy of the spherocylinder must be minimized with respect to the aspect ratio $h/\rho$ while maintaining a fixed area $S = 4\pi\rho^2 + 2\pi\rho h$. For a long spherocylinder, the minimum energy equals $E_{sc}/\kappa = 2\pi + 4\pi\kappa_G/\kappa$ with a cylinder curvature 1/ ρ equal to $C_0$.

By comparing the energy of a sphere with that of a spherocylinder of the same area S, the energy of the sphere is seen to be less than that of the spherocylinder for $C_0R$ less than three while for $C_0R$ greater than three, the spherocylinder has a lower energy. Linear stability analysis[28] shows that the sphere is unstable against small deformations with the symmetry of a spherical harmonic $Y_L^M$ once $C_0R$ exceeds L(L+1), with $L \geq 2$. The sphere thus remains a *local* energy minimum up to $C_0R$ equal to six, where it becomes unstable against an infinitesimal prolate deformation. It follows that we should expect a first-order sphere-to-spherocylinder shape transition for $C_0R \approx 3$.

We also must compare the bending energy $2\pi\kappa + 4\pi\kappa_G$ of a single long spherocylinder, with curvature 1/ ρ equal to $C_0$, to that of a certain number M of spheres with the same total area. We now can set the mean curvature 2/R of the spheres equal to the preferred curvature $C_0$ so the total bending energy of the spheres equals $4\pi\kappa_G M$ (for large M). It follows that – for any value of $C_0$ – a long spherocylinder is stable against break-up into spheres as long as the Gauss curvature constant is positive or, more precisely, as long as $\kappa_G > \kappa/2(M-1)$) with M a large number.

**B) <u>The Lobkovsky Limit</u>**

We now turn to the limit μ = ∞ and λ = ∞, and hence infinite Area *and* Young's Moduli. The sheet is now inextensible, which corresponds to the isometric regime discussed in the previous section where we showed how to construct faceted isometric shells. We would now like to compare the elastic energy of different isometric shells. Because the curvature of an isometric shell is infinite along each of the ridges connecting adjacent facets of an isometric shell, the bending energy of an isometric shell is infinite if Y is infinite. In the limit of large but finite Y, and hence of finite γ, the elastic energy E(L) of a ridge of length L that connects two facets whose normals make an angle 2β with respect to one another was obtained (for zero spontaneous curvature) by Lobkovsky[29] by the use of scaling arguments:

$$E(L)/\kappa \propto \beta^{7/3}\left(\frac{YL^2}{\kappa}\right)^{1/6}. \qquad (3.5)$$

The dimensionless ratio $\gamma(L) = \frac{YL^2}{\kappa}$ of stretching and bending energies can be viewed here as a FvK Number, which must be large compared to one in order for Eq. 3.5 to hold. LMN found that Eq. 3.5 indeed gives the elastic energy of the edge of an icosahedron, provided $\gamma$ exceeds a number of the order of $10^9$. Since the transverse curvature 1/R of an edge scales as $\frac{\gamma(L)^{1/6}}{L}$, the contribution to the bending energy coming from the spontaneous curvature term can be neglected for $C_0 L < \gamma(L)^{1/6}$, i.e., again for sufficiently large FvK Numbers.

In order to compare the elastic energies of different shell shapes in this limit, we computed an energy index E defined by

$$E \propto \sum_{i,j} \beta_{i,j}^{7/3} L_{i,j}^{1/3} \qquad (3.6)$$

for different isometric shells of the same area. The summation extends over the ridges connecting pairs of 5-fold vertices of an isometric shell. Note that Eq. 3.6 can be viewed as the effective interaction energy between twelve particles restricted to a closed surface interacting via a pair potential that increases as the 1/3 power of their separation.

Figure 5A shows the E index of an isometric spherocylinder as a function of the ratio m/n of the isometric construction. The m/n ratio, which equals one for an icosahedron, is about 2.5 h/ρ for a long spherocylinder. The vertical axis gives the E index, divided by the value of E for an icosahedron.

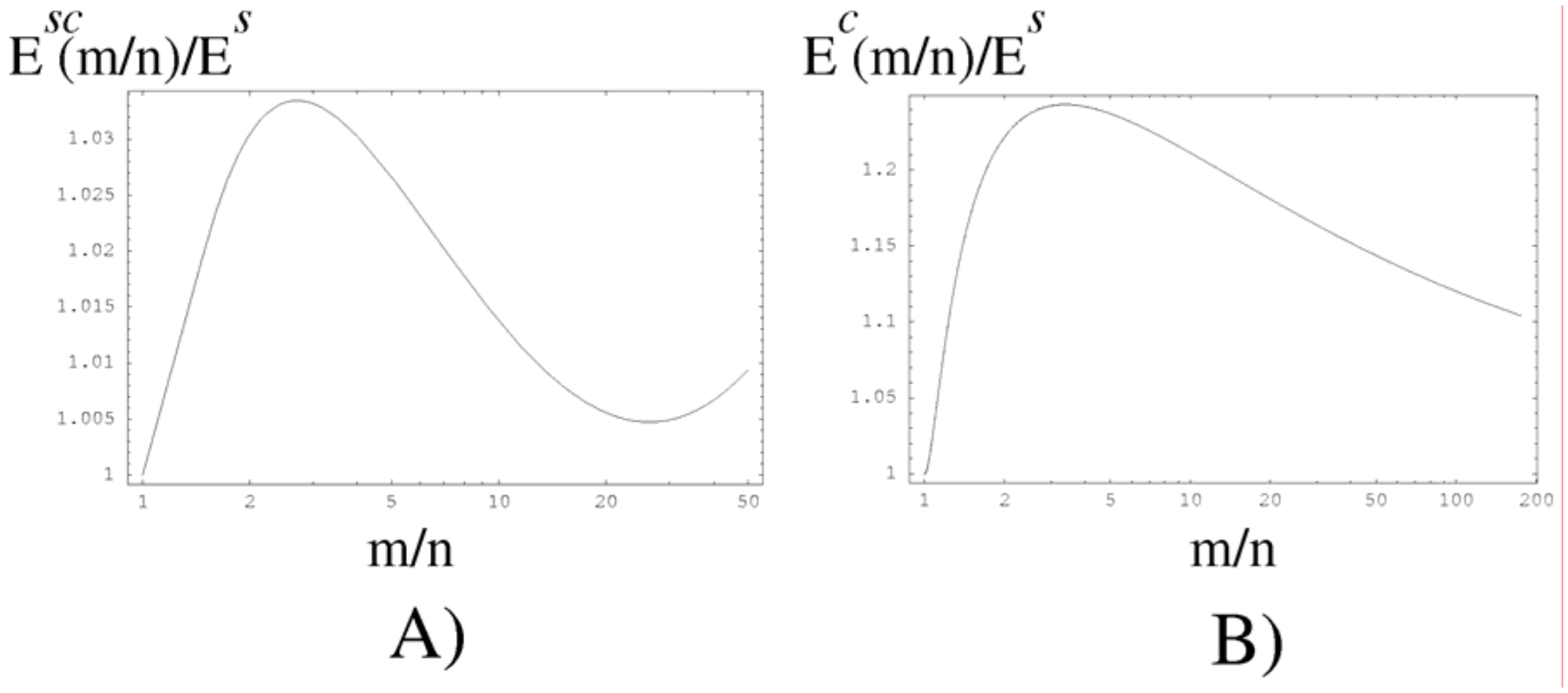

Figure 5. Elastic energy of isometric shells as computed from the Lobkovsky scaling relation Eq. 3.6. The energy is normalized with respect to that of an icosahedral ("s") shell of the same area. Fig. 5A: Elastic energy of a spherocylinder ("sc") as a function of the ratio m/n of the isometric construction. This ratio is about 2.5 times the aspect ratio h/ρ for a long spherocylinder. Fig.5B: Elastic energy of an isometric 5-7 cone ("c") as a function of the m/n ratio of the isometric construction. The m/n ratio is approximately the ratio ($R_l/R_s$) of the radii of the larger and smaller caps of the cone.

The E index at first increases as a function of m/n. The reason is that when we increase the aspect ratio, at fixed total area, we increase the *total* length $\sum_{ij} L_{ij}$ of the ridges of an isometric spherocylinder shell as compared to that of an icosahedral shell. The index develops a maximum but it then decreases for increasing aspect ratio. This decrease is due to the fact that for larger aspect ratios the 12 vertices are densely clustered on the two caps. By decreasing the cap size, we lower the energy of the ten short ridges on each of the two caps much more than we raise the energy of the ten long ridges connecting the caps because of the large negative curvature of the $L_{i,j}^{1/3}$ pair potential for small L values. The energy reaches a minimum when the aspect ratio h/ρ of the spherocylinder is of order ten. At this minimum, the energy of the spherocylinder is very nearly degenerate with that of the icosahedron but the energy of the spherocylinder never drops below that of the icosahedron. Inclusion of spontaneous curvature, which favors cylindrical structures, should thus lead to a first-order, i.e., discontinuous, transition between the icosahedron and a spherocylinder with an aspect ratio h/ρ of order 10, i.e., near the minimum of Fig.5A. It is questionable though whether the scaling description Eq. 3.6 can be extended to spherocylinders with such very large aspect ratios. The angle between two neighboring ridges joining at a vertex is small in this regime, causing overlap of the regions of stress along the ridge near a vertex.

Turning to cones, Fig. 5B shows the normalized E index of an isometric 5-7 cone now as a function of the m/n ratio of the isometric construction, which is approximately the ratio of the radii of the two caps. Note that the elastic energy of the cone is systematically higher energy than that of the spherocylinder so cone-like structures are not expected to be stable in the isometric limit.

**C) Generalized LMN Theory**

Consider the "self-energy" of a single 5-fold disclination defect at the center of a circular sheet of hexagonal material with radius R, a problem that was studied by Seung and Nelson (SN)[30]. If the sheet is forced to remain flat, then the stretching energy of a disclination, computed from Eq. 3.1, diverges as the *area* of the sheet: $E(R) \cong AYR^2$ with A = π/288. If, on the other hand, the sheet is allowed to buckle out of the plane, then it can reduce the elastic energy by forming a *cone* with only a central core region that is flattened out in order to avoid a divergence of the bending energy. The bending energy of the cone section can be easily computed from Eq. 3.2 and is equal to $E(R) \cong B\kappa \ln\left(\frac{R}{R_B}\right)$ with B equal to π (11/30) and with $R_B$ the buckling radius mentioned in the Introduction. The flattened core has a radius of the order of $R_B$ and an energy of order $E_C \cong AYR_B{}^2$. These two results can be combined into a single variational expression for the energy of a disclination in a lattice of size R (>$R_B$):

$$E(R) = AYR_B^2 + B\kappa \ln(\frac{R}{R_B}). \qquad (3.7)$$

Minimizing (3.7) with respect to $R_B$ gives a buckling radius $R_B \approx \sqrt{(B/2A)\kappa / Y}$ **,** allowing (3.7) for the disclination energy E(R) to be written in terms of B and $R_B$ instead of B, $R_B$ and A. More explicitly, we have $E(R > R_B) = \frac{B\kappa}{2}(1 + 2\ln\frac{R}{R_B})$, from which it is clear that the disclination energy increases (decreases) with decreasing (increasing) $R_B$, the size at which the sheet buckles. In terms of $\gamma = YS/\kappa = Y4\pi R^2/\kappa$, this corresponds to a critical FvK number $\gamma_B$ equal to $2\pi$ (B/A) or about 660. A more detailed calculation – for a single disclination in a planar lattice, treating the join between flat and cone portions more carefully via a direct numerical evalutation of the lattice energy – gives a larger value, about 1935 (LMN, SN).

A simple and appealing Ansatz for the elastic energy of a spherical capsid is to add the energy of twelve such disclinations to a background elastic cost for forming a spherical capsid from an equivalent area of planar lattice. Using numerical minimization of the energy, LMN found[8] that this procedure works well for icosahedral capsids up to quite high FvK Numbers, provided one treats B and $\gamma_B$ (and hence $A = 2\pi B/\gamma_B$) as fitting parameters. B had to be increased only slightly (to a value of 1.30 from the approximate value $\pi$ (11/30) $\approx$ 1.15) whereas $\gamma_B$ had to be reduced significantly (from about 1935 to about 1633). This means, according to our discussion following Eq.3.7, that the elastic energy of a disclination imbedded in a spherical shell is greater than the elastic energy of a disclination imbedded in an asymptotically flat sheet. LMN found that, at the disclination buckling transition, the global shape of the capsid undergoes a transition from spherical to polyhedral (see Fig.2). Note that in this approach disclinations effectively interact via a logarithmic pair potential – above the buckling threshold – which again has a negative curvature, as in the Lobkovsky regime.

In order to include the effect of spontaneous curvature in the approach of Nelson and coworkers, we express it as a dimensionless number, $\alpha = C_0 S^{1/2}$, which will form a second important dimensionless variable in addition to the FvK number $\gamma$. For small capsids, with $\gamma$ less than $\gamma_B$, the total continuum elastic energy $E_0$ ($\gamma$, $\alpha$) of a spherical shell with spontaneous curvature equals:

$$E_0(\gamma,\alpha)/\kappa \cong 6B\frac{\gamma}{\gamma_B} + D(\alpha) \qquad (\gamma < \gamma_B)\,. \tag{3.8}$$

The first term is the elastic stretching energy of 12 "unbuckled" disclinations. The background elastic energy D(α) is the quadratic function of spontaneous curvature given by Eq. 3.3, expressed here in terms of the dimensionless α:

$$D(\alpha) = D(0) - 4\sqrt{\pi}\alpha + \frac{1}{2}\alpha^2\,, \tag{3.9}$$

with $D(0) = 4\pi(2 + \kappa_G/\kappa)$. This is just the bending energy, in units of κ, needed to form a defect-free spherical surface from a planar one of the the same area; Eq. 3.8 adds to this quantity the strain energies associated with the 12 unbuckled (flat!) defects that comprise this area.

For capsids with γ greater than $\gamma_B$, the elastic energy $E_0$ (γ, α) equals:

$$E_o(\gamma,\alpha) = 6B(1 + \ln\frac{\gamma}{\gamma_b}) + C(\gamma,\alpha) + \hat{D}(\gamma,\alpha) \qquad (\gamma > \gamma_B). \tag{3.10}$$

The first term equals twelve times our earlier sum – see Eq. (3.7) and discussion following it – of the elastic stretching energy of the flat core section of a disclination plus a bending energy for the curved conical section. The second term, $C(\gamma,\alpha)$, is the contribution of the spontaneous curvature to the bending energy of the conical sections of these disclinations:

$$C(\gamma,\alpha) = 6B\{-\frac{2\alpha}{\sqrt{\pi}}[\sqrt{F(\gamma)} - \sqrt{\frac{\gamma_B}{\gamma}}] + \frac{\alpha^2}{4\pi}[F(\gamma) - \frac{\gamma_B}{\gamma}]\}\,. \tag{3.11}$$

Here $F(\gamma)=\dfrac{1-\frac{\gamma_B}{\gamma}\left(1-3\cos\theta_1/\tan\theta_1\right)}{3\cos\theta_1/\tan\theta_1}$, with $\theta_1$ equal to (half of) the largest cone angle consistent with forming a truncated cone from a hexagonal lattice, i.e., the M=1 case of the truncated cones considered explicitly at the start of the following section – see Eq. 3.13; $\theta_1=\sin^{-1}(5/6)$. Note that $C(\gamma,\alpha)$ is a quadratic function of the spontaneous curvature that vanishes when the cone area goes to zero at $\gamma$ equal to $\gamma_B$. Finally, the third term of Eq.3.10, $\hat{D}(\gamma,\alpha)$, accounts both for the bending and spontaneous curvature energies of the core sections of the disclinations. This term depends on the ratio $\gamma_B/\gamma$ as well as on the spontaneous curvature $\alpha$:

$$\hat{D}(\gamma,\alpha)=D(0)-4\sqrt{\pi}\alpha\sqrt{\frac{\gamma_B}{\gamma}}+\frac{1}{2}\alpha^2\frac{\gamma_B}{\gamma}. \qquad (3.12)$$

Note that for $\gamma=\gamma_B$, $\hat{D}(\gamma,\alpha)$ reduces to $D(\alpha)$ – see Eq. 3.9, since in that case there are no buckled regions and the whole area of the capsid is associated with flat cores that have been bent into a sphere. Otherwise, i.e., for $\gamma>\gamma_B$, the cores comprise only part of this area (specifically the fraction $\gamma_B/\gamma$) and the "background" bending/spontaneous curvature energies are given by $\hat{D}(\gamma,\alpha)$ instead of $D(\alpha)$.

In Fig.6 we show the dependence of the elastic energy $E(\gamma)$ of an icosahedral shell on the FvK Number $\gamma$ for different values of the spontaneous curvature. The spontaneous curvature - rather than the $\alpha$ parameter - was kept fixed here in order to the display the dependence of the elastic energy on system size S.

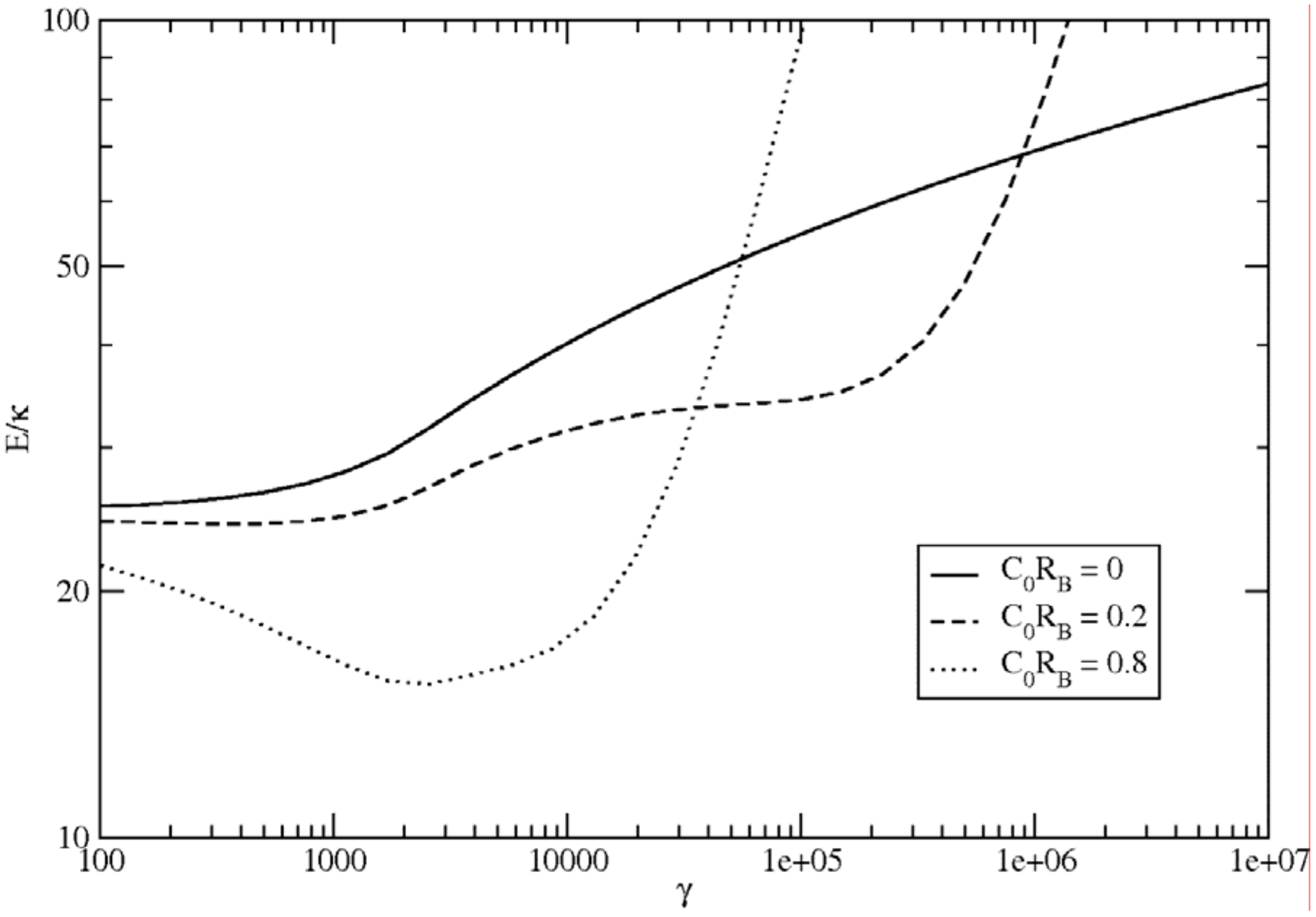

Figure 6: Elastic energy of an icosahedral shell as a function of the Föppl-von Kármán Number $\gamma$ for different values of the spontaneous curvature $C_0$ as predicted by Eqs. 3.8 - 3.12. Solid line: $C_0 = 0$. Above the buckling threshold, the energy has a negative second derivative with respect to area. In this case, the capsid size distribution would be highly polydisperse in a self-assembly experiment. Dashed lined: $C_0R_B = 0.2$. The region of negative second derivative is reduced to a finite interval. Dotted line: $C_0R_B = 0.8$. The energy has a positive second derivative with respect to area. In a self-assembly experiment, capsids with Fvk Numbers near this minimum would dominate.

If the spontaneous curvature is small compared to $\approx 0.1/R_B$, with $R_B$ the buckling radius, then the $E(\gamma)$ curve deviates little from the case of zero spontaneous curvature (solid line) at least for FvK Numbers less than $10^6$. However, if $C_0R_B$ equals 0.2 (dashed line), then the interval of negative curvature of the $E(\gamma)$ curve has substantially diminished. If $C_0R_B$ equals 0.8 (dotted line), then the curvature of $E(\gamma)$ is positive everywhere, while $E(\gamma)$ exhibits a single, well-defined minimum for capsid radii of the order of $1/ C_0$.

**D) Non-Icosahedral Shapes**

[ We now compare the elastic energy $E_0(\gamma,\alpha)$ of an icosahedral capsid with that of conical and spherocylindrical caspids. In order to describe a cone-like capsid, we first approximate a conical shell by a closed surface consisting of two spherical cap portions connected by a smooth cone with an aperture angle 2θ (see Figure 7), and then add the disclinations.

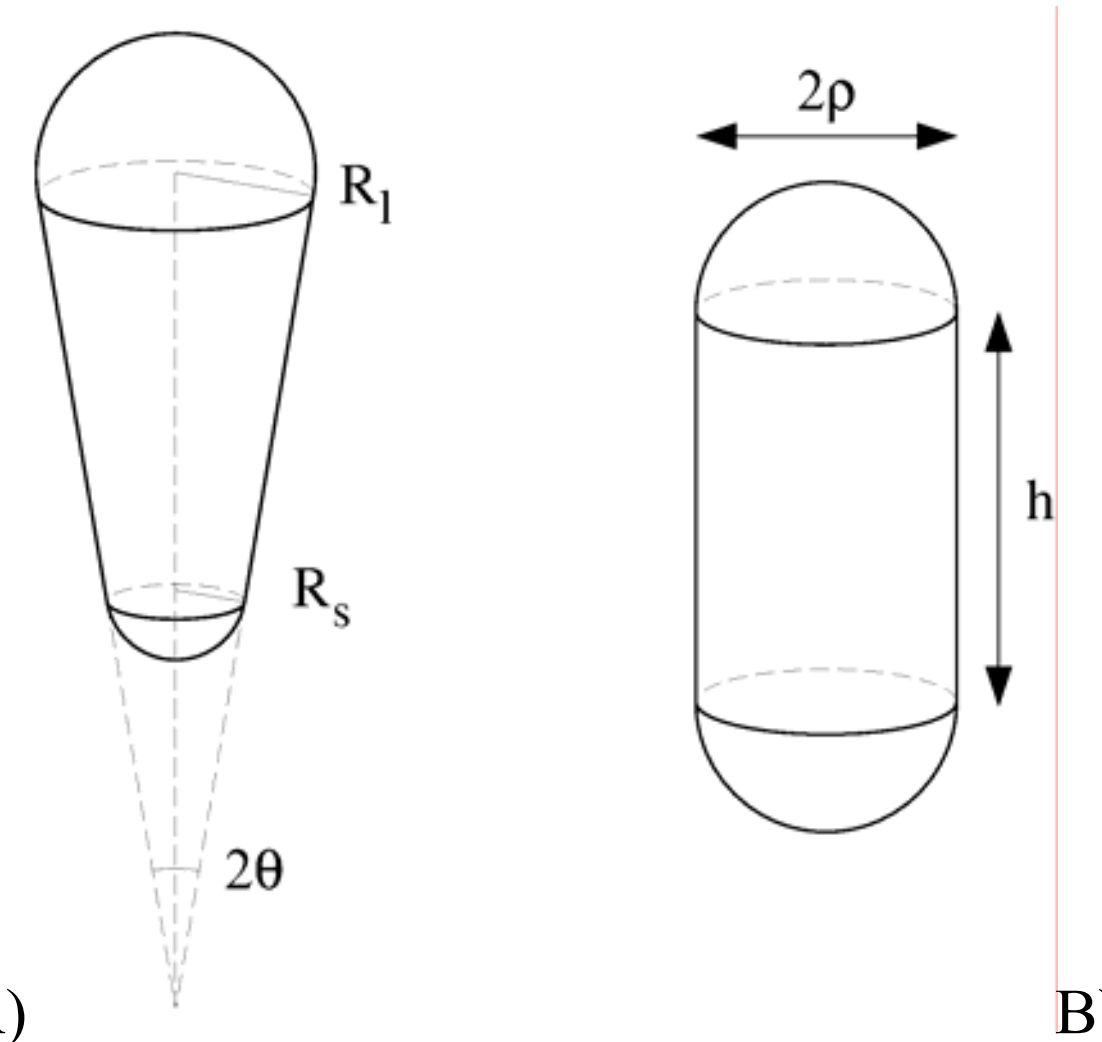

A) B)

Fig.7A: Construction of a smooth conical shell by joining a larger sphere of radius $R_l$ to a smaller sphere of radius $R_S$ by a cone that is cotangent to the two spheres. The cone aperture angle is 2θ. The parts of the spheres inside the cone are then removed leaving two spherical cap portions. Note that the surface has a discontinuity in the curvature along the two matching circles. Fig.7B: Spherocylinder of height h and cylinder radius ρ.

The curvature radius of the larger (top) cap will be denoted by $R_l$ and the radius of the smaller (bottom) cap by $R_S$. We now assign different numbers of disclinations to the two caps: the top cap contains 12-M disclinations and the bottom cap M disclinations. The special M = 6 case corresponds to the spherocylinder. The value of M actually determines the aperture angle of the cone in Fig.7A. The reason is that a cone of aperture angle θ can be constructed from a flat circular sheet by cutting out a wedge with an angle equal to $2\pi(1-\sin\theta)$ and then closing the cut. On the other, introducing a single five-fold disclination into a sheet corresponds to removing a wedge with an angle of 2π/6 from the sheet, and then closing the sheet, which produces a cone with an aperture angle of arcsin 1/6. The aperture angle of the cone is, in general, quantized by the number M of disclinations of the bottom cap:

$$\sin\theta_M = 1 - \frac{M}{6}. \tag{3.13}$$

A conical capsid is thus characterized by *two* FvK Numbers, $\gamma_l$ and $\gamma_s$, for the larger and smaller, respectively, of the two caps: $\gamma_{l,s} = \frac{YS_{l,s}}{\kappa}$ with $S_l = \frac{12-M}{12}4\pi R_l^2$ and $S_s = \frac{M}{12}4\pi R_s^2$. The size of the conical section is fixed once we have specified the two FvK Numbers plus the aperture angle.

The elastic energy $E_M(\gamma_l,\gamma_s,\alpha)$ of the capsid is now approximated as the sum of three terms: the two elastic energies of the caps, computed as the energy $E_0$ of an icosahedral shell but scaled by the appropriate number of disclinations, plus the elastic energy of the connecting cone section:

$$E_M(\gamma_l,\gamma_s,\alpha)/\kappa = \frac{12-M}{12}E_0(\gamma_l,\alpha) + \frac{M}{12}E_0(\gamma_s,\alpha) + D_M(\gamma_l,\gamma_s,\alpha). \tag{3.14}$$

The energy of the cone section is the sum of a bending energy and a spontaneous curvature term that is similar to that of cone section for single disclinations, given by Eq.3.11:

$$D_M(\gamma_l,\gamma_s,\alpha) = \left(\frac{\cos\theta_M}{2\tan\theta_M}\right)\left\{\pi\ln\frac{\gamma_l}{\gamma_s} - 2\sqrt{\pi}\alpha\left(\sqrt{\frac{\gamma_l}{\gamma}} - \sqrt{\frac{\gamma_s}{\gamma}}\right) + \frac{\alpha^2}{4}\left(\frac{\gamma_l}{\gamma} - \frac{\gamma_s}{\gamma}\right)\right\}. \tag{3.15}$$

Here, γ is the FvK Number of a sphere having the same area as the conical capsid, defined again as YS/κ. The special case M = 6, the spherocylinder (see Fig.7B), is more conveniently expressed as:

$$D_6(\gamma_c, \alpha, h/\rho) = h/\rho\left\{\pi - \sqrt{\pi\left(\frac{\gamma_c}{\gamma}\right)\alpha + \frac{1}{4}\left(\frac{\gamma_c}{\gamma}\right)\alpha^2}\right\} \quad (3.16)$$

with $\gamma_c$ the FvK Number of the two caps and with h/ρ the ratio of the height and radius of the cylindrical section.

We now can plot a shape phase-diagram as a function of the parameters γ and α, comparing different shell shapes with the same total area. For a given M value the size ratio $\frac{\gamma_l}{\gamma_s}$ of the top and bottom cap is treated as a variational quantity whose value is determined by minimization of the energy at fixed area. The resulting γ−α shape phase diagram is quite simple, as shown in Fig.8:

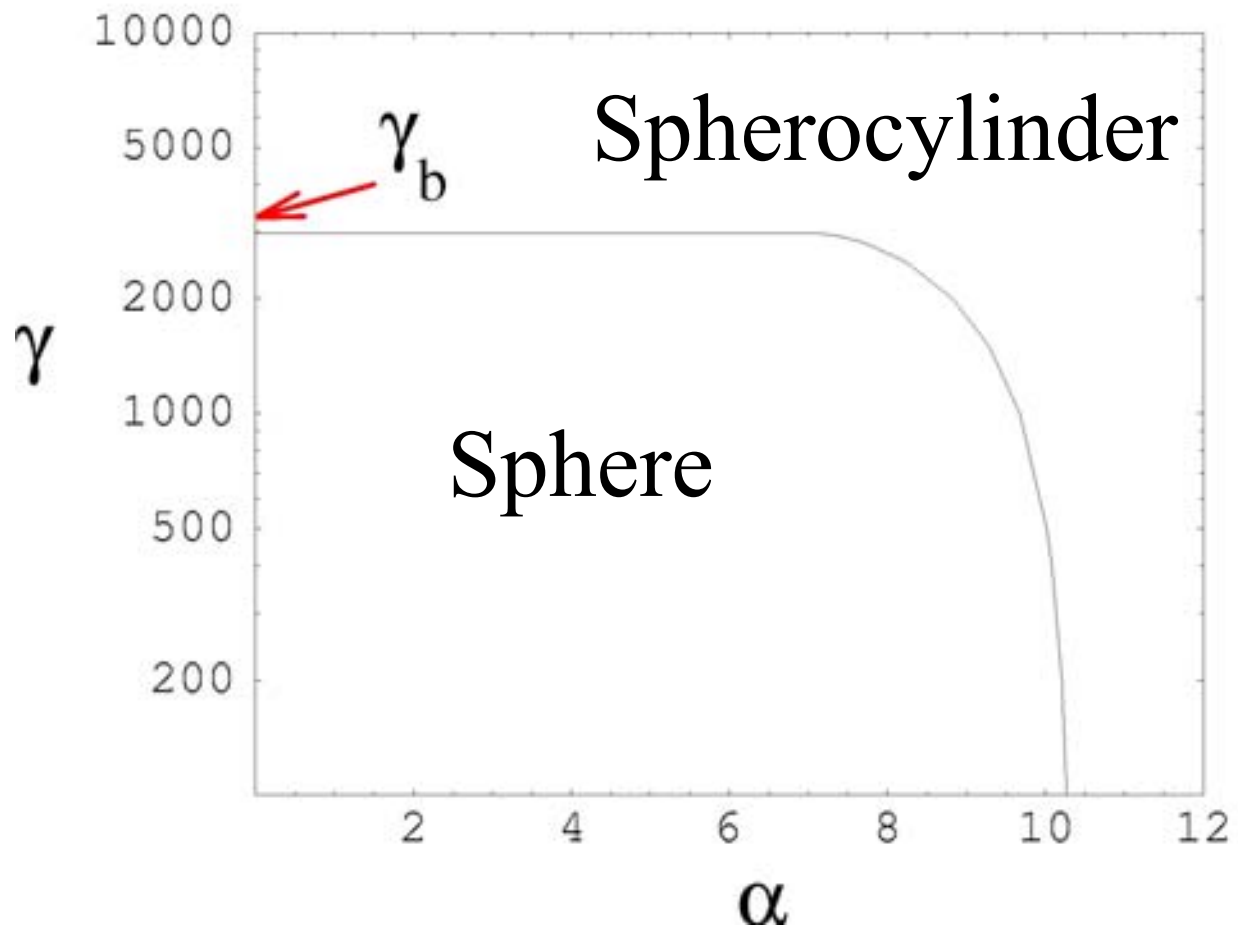

Figure 8: Analytical shape phase-diagram. The vertical axis is the Föppl-von Kármán Number γ = YS/κ, with S the area of the shell. The horizontal axis is $\alpha = C_0 S^{1/2}$, the spontaneous curvature $C_0$ in dimensionless units. Solid line: transition from a spherical shell to a spherocylindrical shell. The transition is discontinuous, with the aspect ratio of the spherocylinder along the transition line ranging from 4.6 to 5.8. Conical shells with $M \neq 6$ do not arise. For small FvK Numbers, the transition takes place close to the critical spontaneous curvature of the Helfrich theory (see Section IIIA). For small values of $\alpha$, the transition takes place, as a function of $\gamma$, just below the numerically computed buckling threshold of the icosahedron of about 3,269 (see Section IV).

For small FvK Numbers ($\gamma < 10^2$), we encounter a first-order transition from sphere to spherocylinder as a function of spontaneous curvature for $\alpha$ near 10.2. A transition point of $\alpha$ near 10.2 translates to a value for $C_0R$ near 3, close to that of the Helfrich limit. The aspect ratio of 5.8 for the spherocylinder at the transition point is significantly smaller than that appearing in the Lobkovsky limit (see Fig.5A). For larger FvK Numbers ($\gamma \approx 10^3$), the transition becomes less dependent on the value of the spontaneous curvature while the aspect ratio slightly decreases. In fact, a transition from sphere to spherocylinder takes place even *for zero spontaneous curvature*. The associated critical FvK number is close to the buckling threshold (red arrow). If the elastic energy of the cone is compared with that of either the sphere or the spherocylinder, then – as for the Helfrich and Lobkovsky limits – one again finds that conical shells never should be stable. Note that the locus of points for which $C_0R_B$ – rather than $C_0R$ – is fixed consists of a family of parabolas in the $\gamma$–$\alpha$ plane.

If the structural phase-diagram of Figure 8 really were to apply to viral capsids, then this would lead to a rather startling prediction: since the FvK Number increases in proportion to the capsid area S, large spherical capsids would be *intrinsically unstable* against the formation of spherocylinders. If one assumes that the values of the Young's Modulus and the Bending Constant are determined by the basic interactions between protein subunits, and that they are therefore similar for different viruses, then this would imply that there should be a *maximum size* for spherical capsids. Also, the stability of self-assembled conical shells of HIV-I capsid proteins indeed could not be understood within the context of continuum elasticity theory.

There is however reason to be cautious about these conclusions. First, recall (Fig.5) that in the Lobkovsky Limit of large FvK Numbers, spherical capsids *were* stable in the absence of spontaneous curvature – though only barely so – which is in disagreement with Fig.8. Next, recall that when the theoretical E(S) curve for an icosahedral shell is compared with the results of numerical energy minimizations it was found necessary to treat the constants B and $\gamma_B$ as fitting parameters and that the energy of a disclination imbedded in a curved surface was found to exceed that of a disclination imbedded in an (initially) flat sheet. It turns out (see Section IV) that this misfit elastic energy depend**s** on the shape of the shell. The conical sections of the five disclinations along the edge of one of the caps of a spherocylinder are in fact more deformed than the conical sections of an icosahedral shell, which could alter the buckling threshold parameter $\gamma_B$. The phase diagram Fig.8, and in particular the presence of a sphere-to-spherocylinder transition for $\alpha$ equal to zero, is sensitively dependent on the assumed value of the buckling threshold value $\gamma_B$.

Finally, our construction of the spherocylindrical shell involved a discontinuity of the curvature along the matching circles between caps and body of the shell (see Fig.7). A curvature discontinuity of an elastic shell is possible only if an external torque is applied to the shell surface. In the absence of such a torque, we must expect a spherocylinder shell to warp in some way to remove the discontinuity and, indeed, this is what we find below in our numerical evaluations of the shell energies, e.g., the spherocylinders develop a “waist” (and hence a region of negative Gaussian curvature).

# IV) Numerical Energy Minimization

In order to verify the analytical results of Section III, we carried out a numerical minimization of the elastic energy H of closed shells. Following LMN, the shell surface was discretized by a closed triangular net of fixed connectivity. The sites of the net were six-fold coordinated, except for the twelve sites with five-fold coordination that are required by Euler's Theorem. The location of the five-fold sites was determined by the demand that, in the isometric limit of large FvK Numbers, their position coincided with the generalized CK constructions of Section II.

The in-plane elastic energy $H_S$ of the net is described as the pair-wise sum of harmonic interaction potential between the nearest neighbors i and j of the net:

$$H_S = \frac{\varepsilon}{2}\sum_{ij}\left(\left|\vec{r}_i - \vec{r}_j\right| - a\right)^2 . \tag{4.1}$$

Here, *a* is the equilibrium spacing of the harmonic potential and ε is the spring-constant, related to the 2D Young's Modulus of the shell by $Y = 2\varepsilon/\sqrt{3}$. This equilibrium spacing should *not* be viewed here as some typical spacing between the subunits of a capsid shell but rather as a discretization length for the numerical minimization of the continuum elastic energy Eqs.3.1 and 3.2. Any characteristic distance scale of the continuum theory, such as the buckling radius or the spontaneous curvature radius, thus should be large compared to *a*.

The out-of-plane bending energy $H_B$ of the net is given as a pair-wise interaction between the normal directions of all adjacent triangles I and J of the triangular net. In the absence of spontaneous curvature, the bending energy is given by:

$$H_B = \frac{k}{2}\sum_{IJ}\left(\hat{n}_I - \hat{n}_J\right)^2 . \tag{4.2a}$$

Here, $\hat{n}_I$ is a unit normal perpendicular to the surface of triangle I. The energy scale k for bending the link between two triangles is related to the Helfrich Modulus by $\kappa = \sqrt{3}k/2$.

In terms of the dihedral angle $\theta_{IJ}$ between the normals $\hat{n}_I$ and $\hat{n}_J$ of the two adjacent triangles I and J, we can write Eq.4.2 as

$$H_B = k\sum_{IJ}\left(1 - \cos\left(\theta_{IJ} - \theta_0\right)\right) \quad (4.2b)$$

We have included in Eq.4.2b the effect of a spontaneous curvature, i.e., the bending energy of two adjacent triangles is minimized by setting the dihedral angle $\theta_{IJ}$ equal to a preferred curvature angle $\theta_0$. If one expands the argument of Eq.4.2b to second order in $\theta_{IJ} - \theta_0$ and evaluates the sum for an infinite cylinder with axis running along a crystal directions and radius large compared to *a*, one obtain an expression of the same form as the Helfrich bending energy (Eq.3.2) with a spontaneous curvature $C_0$ equal to $\frac{k\theta_0}{\kappa a} = \frac{2}{\sqrt{3}}\frac{\theta_0}{a}$.

The elastic energy H was minimized by the conjugate gradient method for closed nets with a large number of sites (typically 30,000). As our starting state, we used the isometric shells of the generalized CK construction described in section II. The reference structure was an icosahedral shell with T = $55^2$ (i.e., h=55, k=0) having 10(T-1)+12 or 30,252 sites (see Eq.2.2). The (minimized) elastic energy of that shell was already shown in Fig.2 as a function of the FvK Number for the case of zero spontaneous curvature. The surface area S in the FvK Number $\gamma = YS/\kappa$ is taken here as the area of the *unstretched* isometric starting structure, and not the actual area. The best fit between the numerical results and Eq.3.7 was obtained for B ≈ 1.27 and a critical FvK Number $\gamma_b$ for the buckling transition of about 3,269 (for reasons that are not clear, this is twice the value reported by LMN of about 1633).

In order to compare the elastic energies of icosahedral, spherocylindrical and conical shells, one must generate nets with the same area (number of sites). However, we saw that the CK construction restricts the number of sites to certain magic numbers so it is, in general, not possible to obtain two nets of different symmetry having exactly the same number of sites. As a first example of a spherocylindrical shell, we used an isometric net defined by m=80, n=38, h=1, and k=0 (see Section II). The corresponding number of sites is 30,402, according to Eq.2.5, so the variation in the shell area δS/S, as compared with the T = $55^2$ icosahedral shell is about 0.5%. Above the buckling threshold, the corresponding overestimate of the energy is of order $\frac{\delta E}{E} \approx \frac{1}{\ln\left(\gamma / \gamma_B\right)} \frac{\delta S}{S}$ using Eq.3.10, so the systematic error in comparing energies of competing structures is of the order of 0.5%. Typical results of the energy minimization for zero spontaneous curvature are shown in Fig. 9, where we compare the elastic energy of the m=80, n=38, h=1, and k=0 spherocylinder (thick line) with that of the (55, 0) icosahedron (thin line).

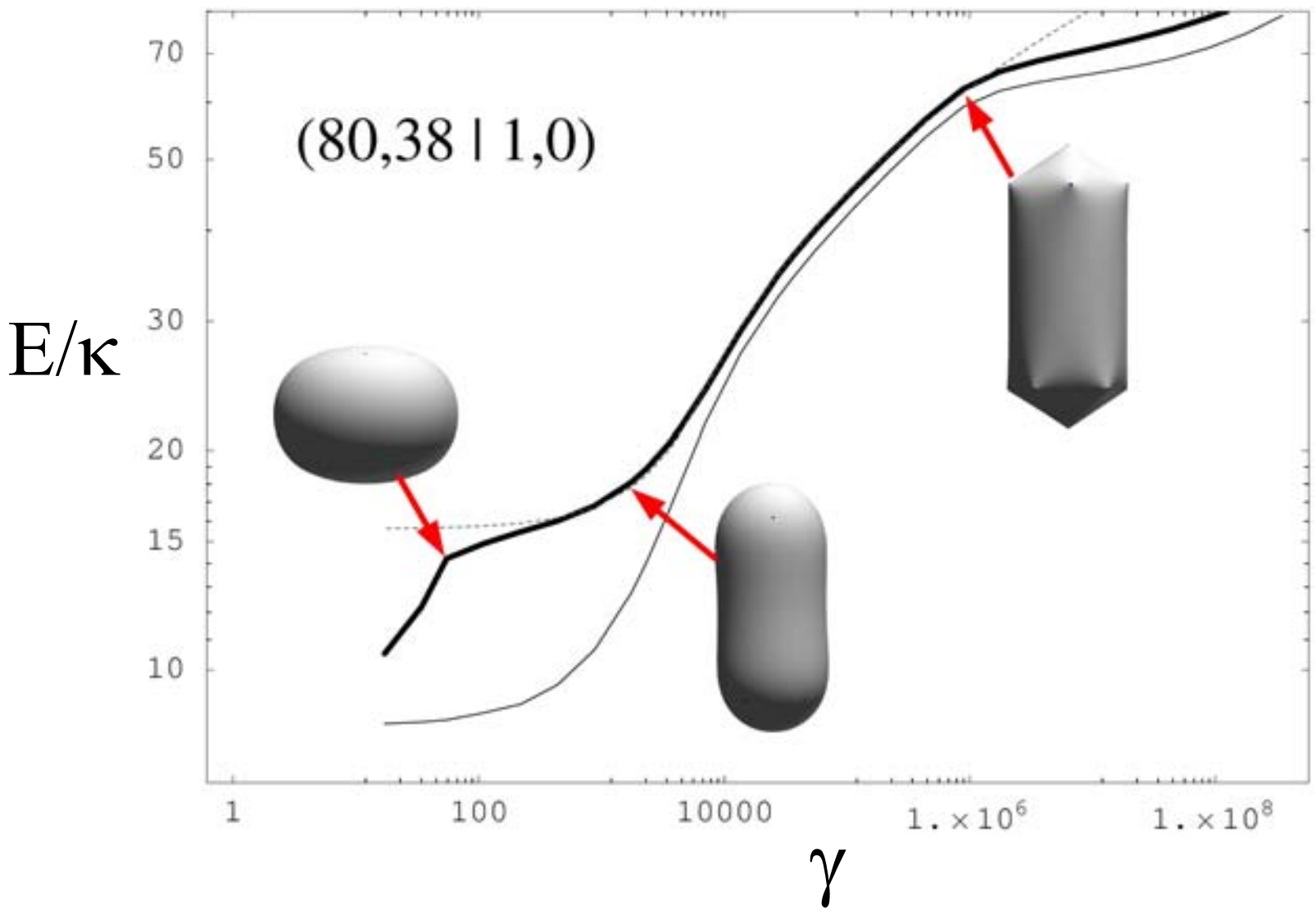

Figure 9: Elastic energy E of a spherocylindrical shell with m=80, n=38, h=1, and k=0 (thick line) and an icosahedral shell with h=55 and k=0 (thin line) for the case of zero spontaneous curvature (in units of the bending constant κ) as a function of the FvK Number γ = YS/κ, with S the area of the shell Y the Young's Modulus and κ the bending constant. The central portion of the spherocylinder develops a negative Gauss curvature ("waistlike region") for γ values around 1,000, followed by a buckling transition for γ near $\gamma_B \approx 2{,}967$ that is smaller than the buckling threshold of a spherical shell (≈ 3,269). The dotted line shows the result of a fit to Eqs.3.14 with adjusted values for B and $\gamma_B$ (see text).

For low FvK Numbers, the shell shape is a prolate ellipsoid, which transforms into a "Dumbbell" shape for FvK Numbers of the order of 1,000. Dumbbell shapes are in fact encountered in the shape catalogue of liquid vesicles[26] but there again only if one imposes a fixed volume constraint. For the present case we note that the Dumbbell shape avoids the curvature discontinuity discussed at the end of Section III. A buckling transition of the spherical caps takes place at $\gamma_B \approx 2,967$. For very large FvK Numbers, above $10^6$, we recover the isometric spherocylinder.

The energy of the spherocylindrical shell always exceeds that of the icosahedral shell, though over a substantial range of FvK Numbers the energy difference ΔE is as low as the "background" energy difference between a fluid sphere and spherocylinder of the same area (and modest aspect ratio), i.e., it is of the order of the bending constant κ. Though small, this energy difference still significantly exceeds the estimated systematic error δE/κ. For instance, for $\gamma \approx 10^4$, δE/κ is of order 0.015, while ΔE/κ is of order 2.0.

If we treat the B and $\gamma_B$ constants as fitting parameters, we can obtain a surprisingly good fit between the theory (see Eq.3.14, dotted line) and our numerical results, except for low FvK Numbers. The fitted buckling threshold $\gamma_B \approx 2,927$ is significantly below the buckling threshold of an icosahedron, though the fitted value of B remains in the range 1.27 - 1.30. Since the self-energy is of the form of $B\left(1+\ln\left(\frac{\gamma}{\gamma_B}\right)\right)$, this indicates that the misfit energy of a disclination that is part of a spherocylinder is larger than the misfit energy of a disclination that is part of a spherical surface with the same FvK number.

The reason for the deviation between numerical energy minimization and the theoretical fit in Fig. 9 for very low FvK Numbers is that the bending energy dominates in that regime over the in-plane elastic energy. As a result, the spherocylinder is deformed towards a spherical shape since the sphere – but not the spherocylinder – is a minimum of the bending energy in the absence of spontaneous curvature.

We repeated this calculation for different values of m and n, while maintaining a fixed area within 0.5%. For general m and n, the elastic energy of the spherocylinder always exceeds that of the sphere, as in Fig.9, but in contrast to the analytical phase diagram of Fig.8. The stabilization of the sphere is precisely because of the increased disclination misfit energy of the spherocylinder, which apparently must be considered as an important physical ingredient in the shape phase diagram of shells.

We have obtained the dependence of the buckling threshold $\gamma_B$ (m,n) on the aspect ratio of the spherocylinder, which is proportional to m/n. Naively, one would expect the caps of a spherocylinder to buckle when the effective FvK Number of a cap, $(1/2)YS_{cap}/\kappa$, is of order the buckling threshold $\gamma_B$ of an icosahedral shell (with $S_{cap}$ the surface area of one of the caps). Since the ratio $S/S_{cap}$ for a spherocylinder equals $(1+h/2\rho)$, with $h/2\rho$ the aspect ratio, one would expect the buckling threshold to be a linearly increasing function of the aspect ratio. The result of a fit of $\gamma_B$ (m,n) as a function of the m/n ratio is shown in Fig.10:

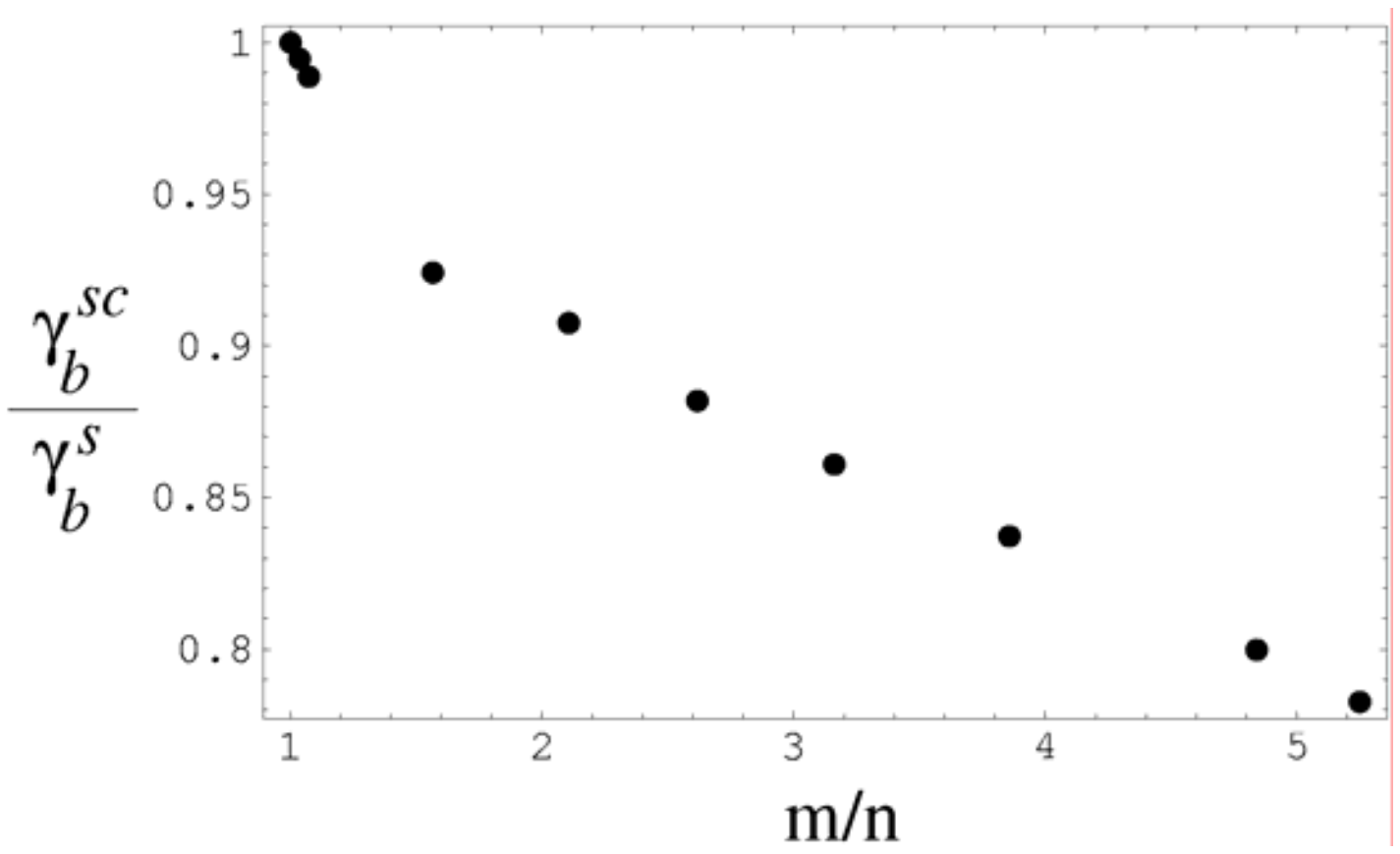

Figure 10: Dependence of the buckling threshold parameter $\gamma_B$ of a spherocylinder (“sc”) – relative to that for a sphere (“s”) – on the m/n ratio, which is proportional to the size aspect ratio. The values of $\gamma_B$ were obtained from a fit of Eq.3.14 to the results of numerical energy minimization.

The buckling threshold indeed has a linear dependence on m/n – which is proportional to the aspect ratio – but $\gamma_B$(m,n) in fact *decreases* with increasing aspect ratio. This means that the misfit energy of a disclination increases with the aspect ratio of the spherocylinder.

Next, we compared the energy of the icosahedral shell with that of a 7-5 conical shell with lattice vectors m=42, n=22, h=1, k=0 and with 30,422 sites (see Eq.2.7). The surface area is again about 0.5% larger surface area than that of the icosahedron and the corresponding error in the energy is of the order of 0.1 κ for FvK Numbers in the range of $10^4$.

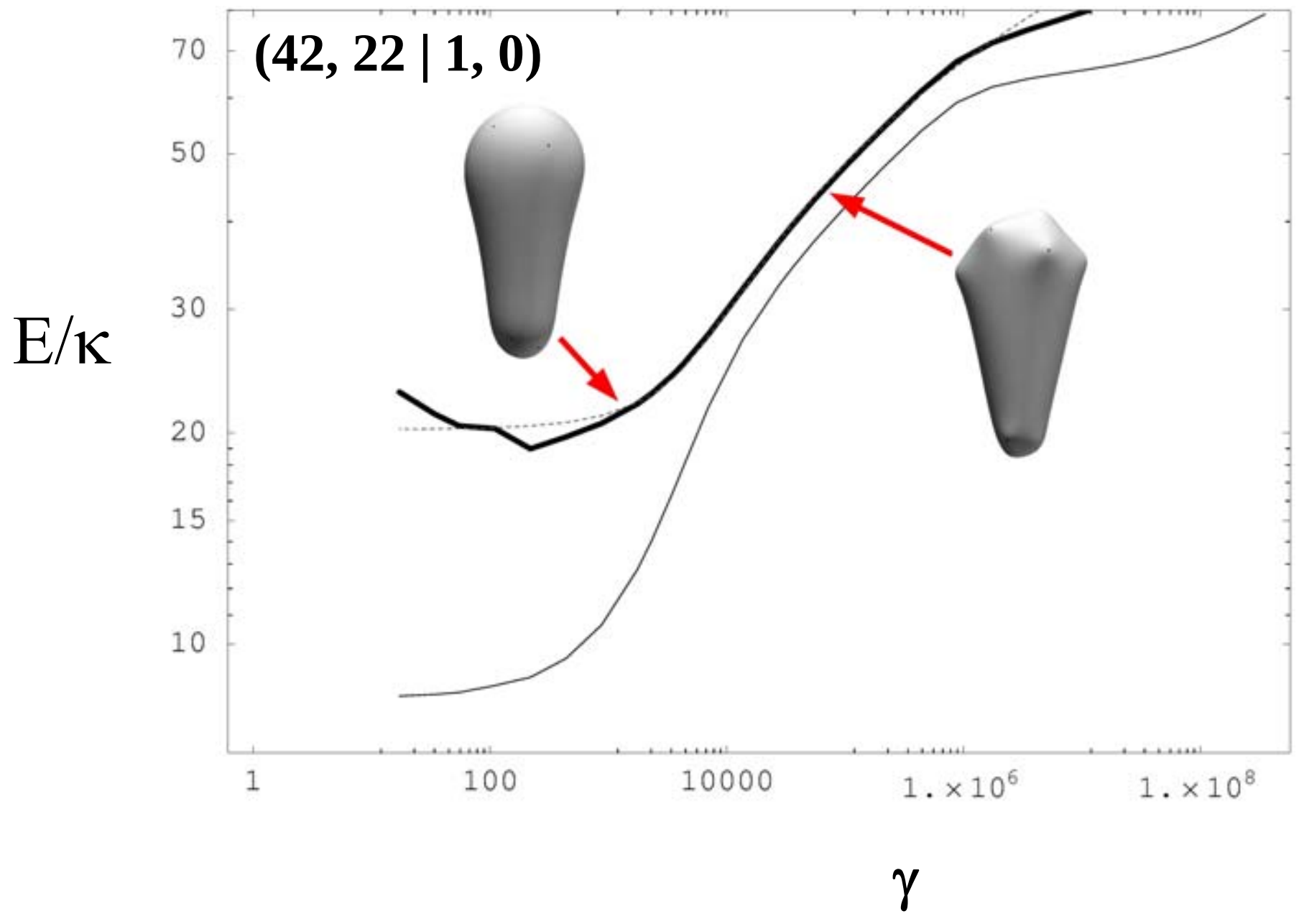

Figure 11: Elastic energy E/κ of a conical shell with m = 42, n = 22, h=1, k=0 (thick line) and the icosahedral shell with h=55 and k=0 (thin line) for the case of zero spontaneous curvature obtained by numerical minimization as a function of the FvK Number $\gamma = YS/\kappa$. Note the pronounced negative Gaussian curvature of the shell. The dotted line shows the result of a fit to Eqs.3.14 with $B \approx 1.36$ and $\gamma_B \approx 4486$.

The elastic energy of the cone is noticeably larger than that of both the spherocylinder and the icosahedron. This is consistent with the analytical results of Section III, where we found that this was due to the extra bending energy of the cone region. It seems that, unlike the spherocylinder, the conical structure does not "compete" with the icosahedral shell, at least in the absence of spontaneous curvature effects. The numerical results can be fitted rather well by Eqs. 3.14 and 3.15 with B ≈ 1.36 and $\gamma_B \approx 4{,}486$. The dependence of the buckling threshold on the cap size-ratio m/n is shown in Fig.12:

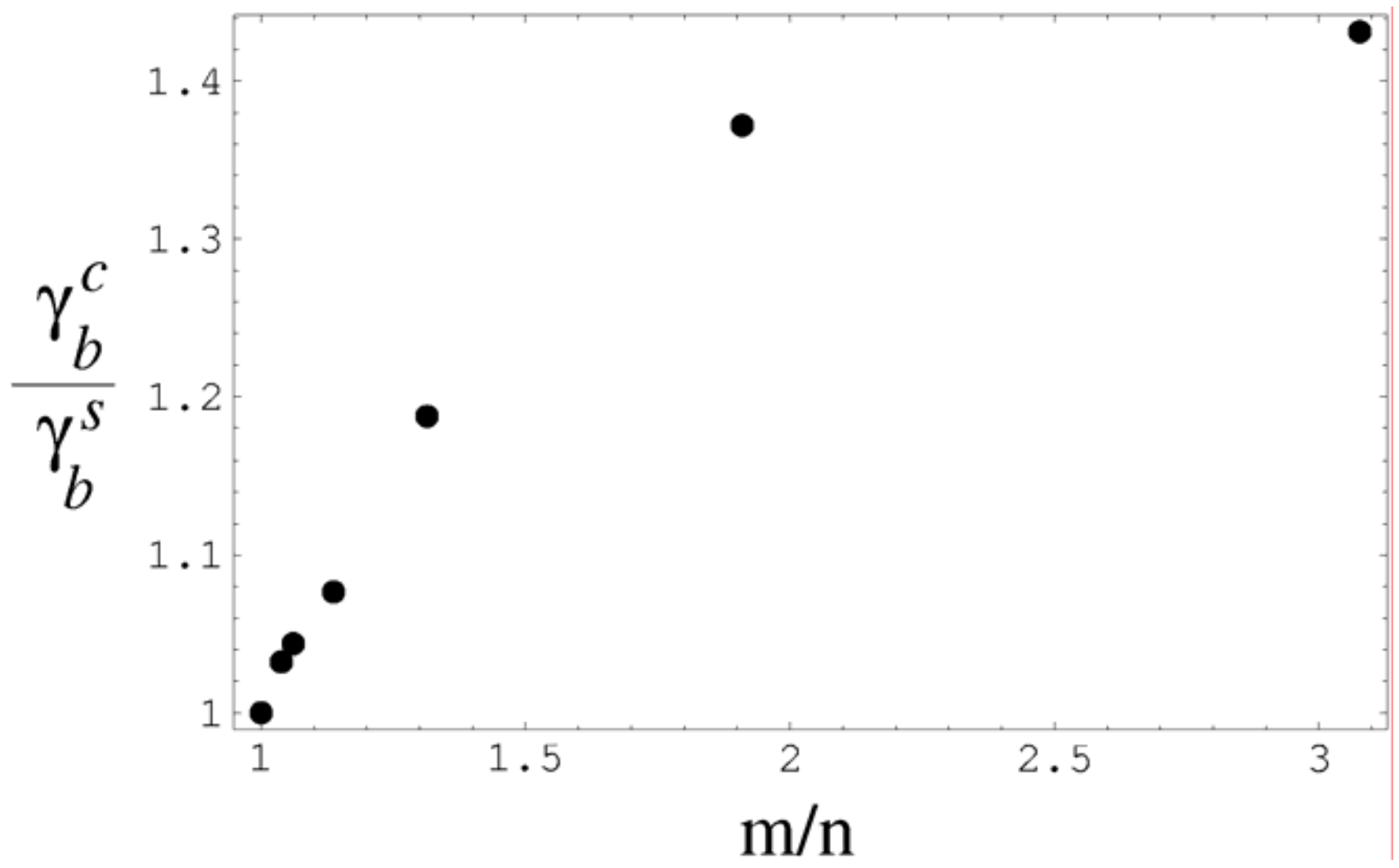

Figure 12: Dependence of the of the buckling threshold value $\gamma_B$ of a conical shell ("c") – relative to that for a sphere ("s") – on the m/n ratio, which is proportional to the cap size-ratio. The values of $\gamma_B$ were obtained from a fit of Eq.3.14 to the results of numerical energy minimization.

In contrast with Fig.11, the buckling threshold now *increases* with the cap size-ratio, which indicates a *decreasing* disclination elastic energy. Even though, over the same range of m/n values, the fitted B coefficient increased from 1.28 to 1.35, the misfit energy indeed does decrease with increasing m/n when assuming the expression $B\left(1+\ln\left(\frac{\gamma}{\gamma_B}\right)\right)$ for the disclination self-energy above the buckling threshold. This leads to the surprising result that the misfit energy appears to *favor* a conical structure.

Next we included the spontaneous curvature energy in order to determine a shape phase-diagram with the FvK Number $\gamma$ and the spontaneous curvature $\alpha$ as coordinates. We determined for given $\gamma$ and $\alpha$ the elastic energy of both the spherocylinder and the cone over a range of m and n values. The m and n values were chosen so that the number of sites always is within 0.5% of that of the T = $55^2$ icosahedral reference structure. We then picked the m and n numbers of the structure that had the lowest energy. The basis vectors of the template were always taken to lie along one of the lattice directions of the hexagonal lattice. Although the spontaneous curvature term significantly alters the *relative energy balance* of the competing shapes, the actual shape of a shell changed by only a minimal amount as compared with the case of zero spontaneous curvature.

The resulting shape phase diagram is shown in Fig.13. The vertical axis is the FvK Number and the horizontal axis $\alpha = \frac{2}{\sqrt{3}}\frac{\theta_0}{a}S^{1/2}$.

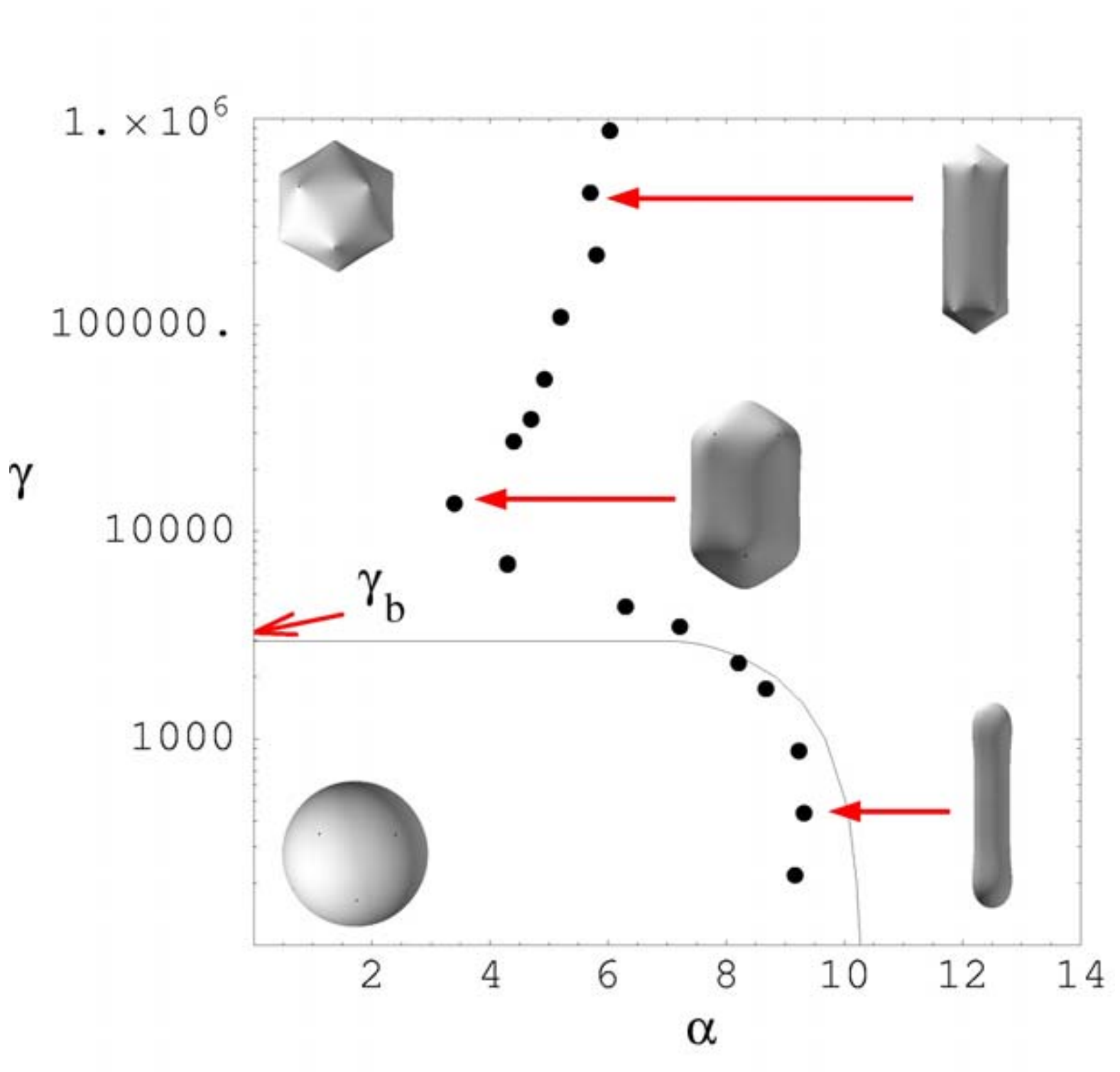

Fig.13: Shape phase-diagram. The vertical axis is the FvK Number $\gamma = \frac{YS}{\kappa}$; the horizontal axis is the dimensionless spontaneous curvature $\alpha = \frac{2}{\sqrt{3}}\frac{\theta_0}{a}S^{1/2}$. For low α, icosahedral shells are stable for all FvK Numbers. The buckling threshold $\gamma_B$ separates spherical from polyhedral shells. For increasing α and FvK numbers below the buckling threshold, a first-order transition line separates spherical and spherocylindrical shells. The m/n ratio of the spherocylindrical shell along the transition line equals 121/25. For increasing α

and FvK numbers above the buckling threshold, the aspect ratio of the spherocylinder at the transition line is reduced and the transition is either weakly first-order or continuous. The solid line shows the phase boundary between sphere and spherocylinder according to the theory described in Section III (see Fig.8).

Only spherical and spherocylindrical shells appear in the phase diagram, as already predicted by the theory of Section III (Fig.8). For low FvK numbers, the transition between these two structures (solid line) takes place reasonably close to the boundary line predicted by the analytical theory (dashed line). The m/n ratio of the spherocylinder at the transition equals 121/25. Figure 14A shows the dependence of the elastic energy on the m/n ratio at the transition point when $\gamma$ equals 873.

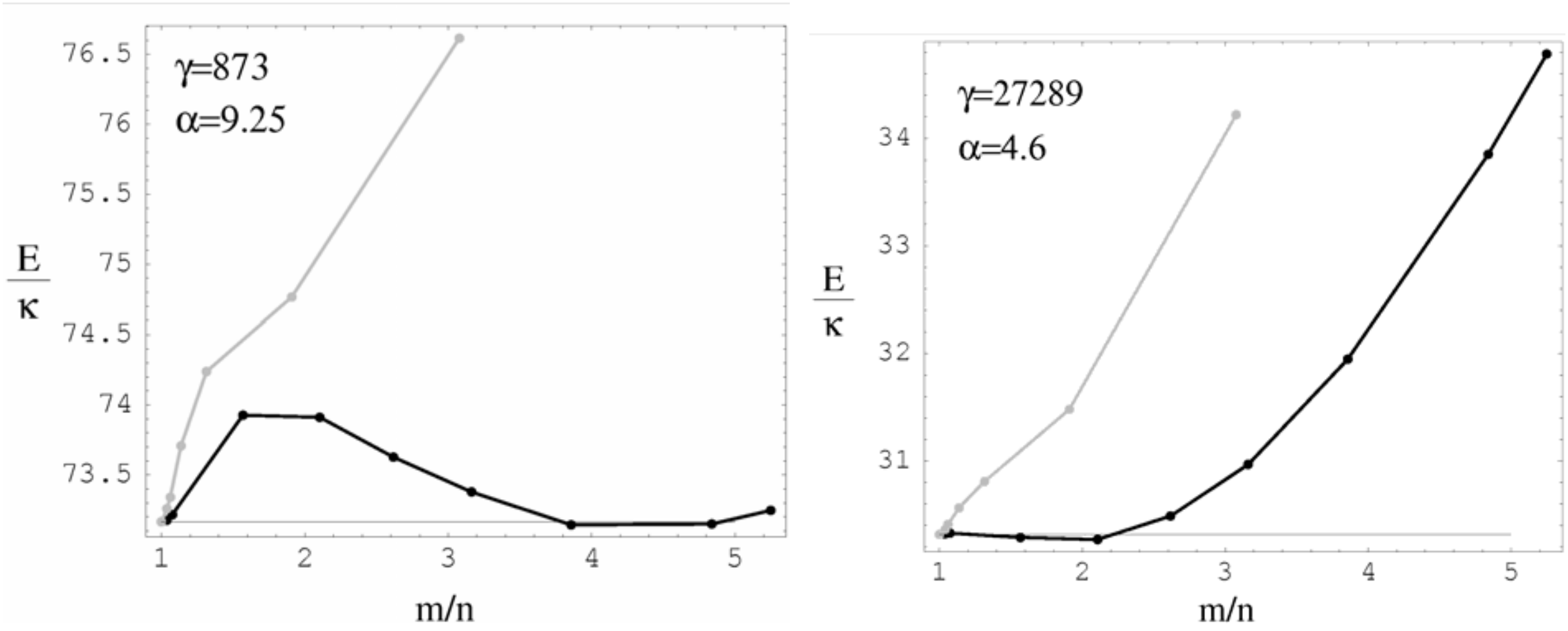

A) B)

Figure 14: Dependence of the elastic energy E, in units of the bending constant κ, of the spherocylinder (black line) and cone (grey line) on the m/n ratio at the transition point. Fig.14A: The FvK Number $\gamma$ = 873 is below the buckling threshold; the energy barrier separating sphere and spherocylinder is of the order of κ. Fig.14B: The Fvk number $\gamma$ = 27289 is above the buckling threshold; the energy barrier is of the order of 0.1κ. The energy of the spherocylinder (black line) is nearly independent of the m/n ratio.

The energy barrier separating the two degenerate structures is of the order of the bending energy constant $\kappa$, as for the $\gamma = 0$ Helfrich theory. The elastic energy of a conical shell for the same values of $\gamma$ and $\alpha$ rises rapidly as a function of the m/n ratio.

When the FvK number approaches the buckling threshold, the value of the critical spontaneous-curvature rapidly shifts to lower values of the spontaneous curvature, as predicted by the analytical theory (see Fig.8), but it never drops below about a third of the maximum value at $\gamma=0$. For low enough values of the spontaneous curvature i.e., $\alpha<3$, the icosahedral shell is in fact stable for any FvK Number. As already noted, the physical origin of the stability of the icosahedral shell can be traced to the lower misfit energy (relative to that in the spherocylinder) of the 5-fold disclination sites.

As we increase the FvK number beyond the buckling threshold, the aspect ratio of the spherocylinder at the transition point is clearly reduced. Figure 14B shows the dependence of the elastic energy of spherocylinder and cone for an FvK Number $\gamma$ equal to 27,289, again at the transition point ($\alpha$=4.6) between spherical and spherocylinderical shells. The energy barrier is significantly reduced as well, to a value of less than $0.1\kappa$. At even higher FvK numbers, the aspect ratio of the spherocylinder starts to increase again (see Fig.13). Recall that in the Lobkovsky scaling limit the transition should be weakly first-order, with an energy barrier of about 3% (see Fig.5a), and a large aspect ratio. Our results suggest that the sphere-to-spherocylinder transition could be *continuous* near the buckling threshold but this cannot be ascertained within our numerical precision of 0.5% of the total energy.

The energy of a conical shell rises rapidly at the transition line as a function of the m/n ratio (see Fig.14B). However, according to the Lobkovsky scaling theory, it is possible that for larger m/n ratios the energy of conical shells should start to decrease (see Fig.5B). We indeed find that this can take place but only at larger values of the spontaneous-curvature. Figure 15 shows the cone and spherocylinder energy, again for $\gamma$ = 27,289 but now at about twice the critical spontaneous curvature ($\alpha$=8).

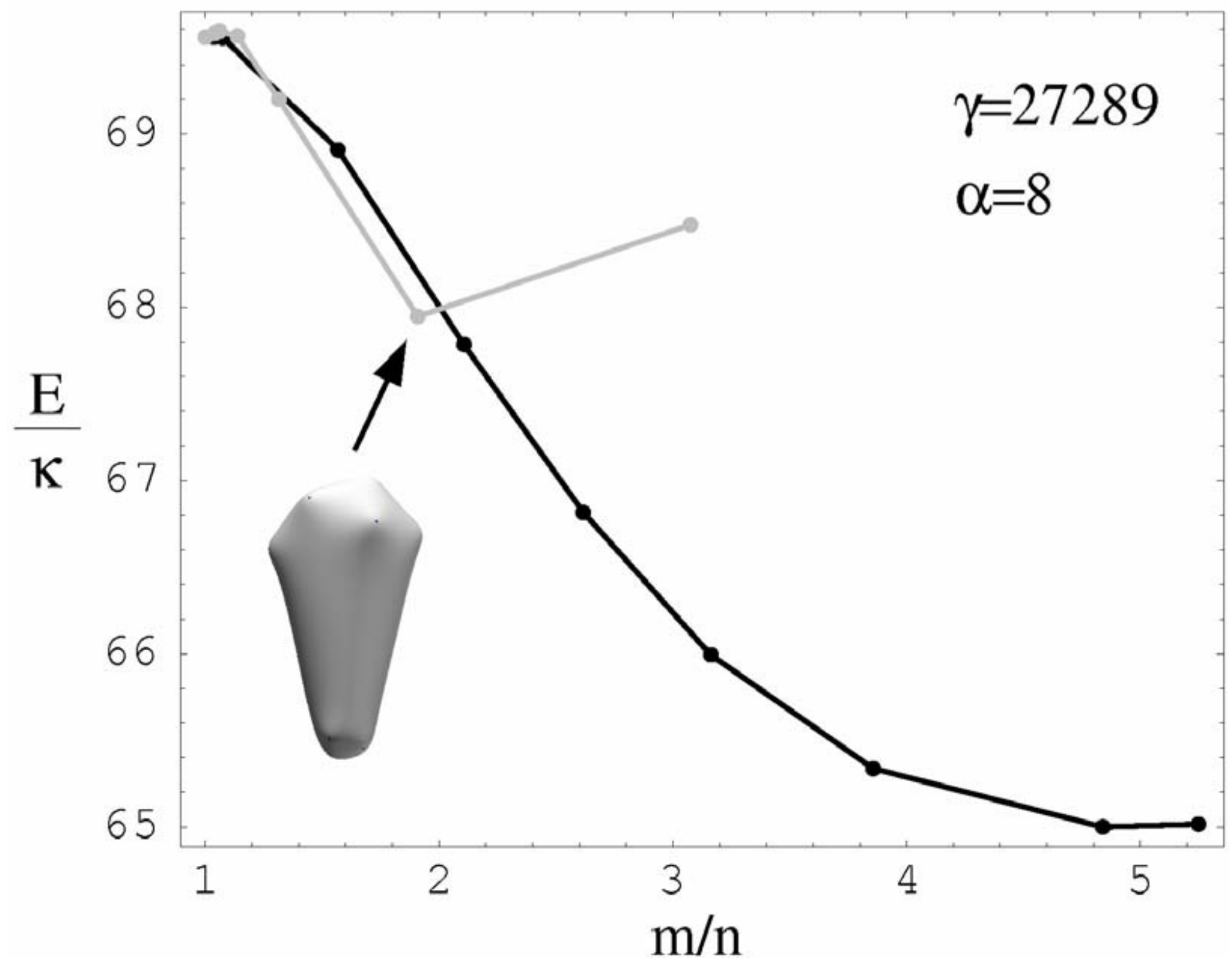

Fig.15. Dependence of the elastic energy of the cone (grey line) and the spherocylinder (black line) on the m/n ratio at about twice the critical spontaneous-curvature and with an FvK number $\gamma$ about twice the buckling threshold. The cone energy has a minimum for m/n=21/11 and the spherocylinder energy has a minimum at m/n=121/25.

The elastic energy of the cone develops a minimum for m/n= 2. The shape of the shell at the minimum is shown in the figure**,** and its energy is lower by about 1.5 $\kappa$ compared with that of the icosahedral shell, though higher by about 3$\kappa$ than that of the spherocylinder at its minimum energy configuration.

# V) Discussion

In this concluding section, we discuss predictions of the continuum theory that can be directly confronted with studies on viral assembly. In order to carry out a comparison it is however critical to know approximate values for the phenomenological constants that enter the theory.

**A) The Young's Modulus and Bending Constant of Viral Capsids**

The FvK Number of a virus can be determined by fitting shell shapes calculated on the basis of the continuum theory to the structure of capsids as determined by X-Ray Crystallography or Cryo-TEM. This procedure was carried out by LMN for the yeast L-A virus, which has a diameter of 43 nm. They obtained a value of $YR^2/\kappa \approx 547$, so the ratio of the Young's Modulus and the bending constant for the L-A virus would be $Y/\kappa \approx 1.24\ nm^{-2}$. They found a similar value of Y/κ (larger only by 30%) from fitting the shape of the – unrelated – bacteriophage HK-97, so we will assume that $Y/\kappa \approx nm^{-2}$ for viral shells in general.

There are at least two different ways to proceed in estimating the individual values of the Young's Modulus and of the bending constant. The first method is by fitting the elastic energy of a spherical shell computed within continuum theory to the results of numerically determined energies of capsid shells. In a recent Monte Carlo simulation of a coarse-grained capsomer model, for example, the total energy of spherical caspids was computed as a function of the number N of capsomers, up to N=80[31]. When one fits the continuum energy (see Fig.2) to the results of that simulation, one obtains a value for κ that is of the order of the capsomer-capsomer cohesive binding energy ε. This cohesive binding energy has been computed in semi-empirical all-atom numerical simulations[32] as well as measured by thermodynamic means[33] for the T=4 Hepatitis B virus. In both cases ε was found to be of the order of 10-15 $k_BT$. It would then follow, from $\kappa \approx \varepsilon$ and the earlier estimate of Y/κ, that the Young's Modulus is about 10 $k_BT$ per $nm^2$.

The second method to estimate Y and κ is by measuring the mechanical deformation of capsids. Atomic Force Microscopy (AFM) studies of two (unrelated) viruses – Φ29[34] (60 nm diameter) and CCMV[35] (30 nm diameter) – report that under an applied load a capsid shell responds like a harmonic spring with a spring constant of about 0.1 N/m. Since the spring constant of an elastic shell is of the order of $\sqrt{\kappa Y}\,/\,R$ according to continuum elasticity theory[36], one can combine the measured value of the spring constant and the earlier value of Y/κ to obtain a value for Y that is of the order of one N/m and a value for κ of the order of $10^{-18}$ J. One also can estimate the *2D* Young's Modulus Y of a protein shell by multiplying its *3D* Young's Modulus (a little less than a GPa, say, as approximated by that of a bulk protein material such as silk) with the typical nanometer thickness of a viral shell (2-3 nm). This again gives about one N/m for Y, while the bending constant – estimated as the 3D Young's Modulus times the cube of the shell thickness – would again be $10^{-18}$ J.

If the estimates for the elastic moduli produced by the second method were valid, then the stored *equilibrium* elastic energy of a large capsid – about 20 κ according to Fig.2 – would be as large as $10^4$ $k_BT$ while the total cohesion energy of a capsid, as measured by thermodynamic means, is actually only of the order of $10^3$ $k_BT$ (again for the Hepatitis B Virus). This would imply that self-assembly of capsid shells was impossible. It is in fact known, from single-molecule studies, that protein-protein interaction forces and energies measured by AFM at finite force loading rates – about 10-100 pN/sec – can be much higher than the actual equilibrium values. For these reasons, we will adopt the estimates of κ and Y of the first method.

**B) Capsid Self-Assembly and Polydispersity**

Here we consider briefly the predictions of the continuum theory in the context of the theory of self-assembly under conditions of thermodynamic equilibrium. The relevance of equilibrium theory for viral assembly can be questioned – fully formed capsids are unlikely to be in equilibrium with a solution of subunits – but it has been shown to be applicable in particular well-studied cases. For example, over thirty years ago, Bancroft and Adoph and Butler demonstrated[15] that capsids of CCMV could be self-assembled from pure protein at low pH, disassembled at high pH, and then re-assembled back at the original low pH.

Under conditions of thermodynamic equilibrium, the concentration c(S) of capsids constructed from S capsomers is given by the Boltzmann distribution $c(S) \propto e^{\left([-\mu_c+\varepsilon]S - E(S)\right)/k_B T}$ with $\mu_c$ the solution chemical potential of the capsomers, ε the cohesion energy per capsomer of an infinite flat hexagonal protein sheet, and E(S) the energy cost of closing the sheet into a shell, as computed for example in the preceding sections. As the chemical potential is increased, capsid shells will start to form when δμ = - $\mu_c$ + ε approaches zero. It follows from the Boltzmann distribution that when two capsids with a different structure – but the same number of subunits – have E(S) values that are within a few $k_B T$ of each other, then we should expect to encounter both structures in a self-assembly experiment carried out under conditions of thermodynamic equilibrium. Self-assembly of a monodisperse solution of capsids requires E(S) to have a well-defined maximum.

For the case of an icosahedral shell with zero spontaneous curvature, we saw in Section III that E(S) is a logarithmic function of S beyond the capsid area $S_B$ of a capsid at the buckling threshold (see Eq.3.5). This means that for larger S the equilibrium profile c(S) should exhibit a *power-law* dependence on S in the absence of spontaneous curvature. The power-law divergence is truncated when S drops below $S_B$:

$$c(S) \propto \begin{cases} \exp^{-(\delta\mu S + 6B\kappa)/k_B T} \left(S_B / S\right)^{6B\kappa/k_B T} & (S > S_B, C_0 = 0) \\ \exp^{-(\delta\mu + AY/4\pi)S/k_B T} & (S < S_B, C_0 = 0) \end{cases}. \tag{5.1}$$

It follows from Eq.5.1 that if the exponent $6B\kappa / k_B T$ of the power-law is large compared to one, then capsid formation – as δμ approaches zero – is restricted to S values of order the buckling threshold $S_B$ or smaller. Note too that below the buckling threshold, the effect of elastic stress simply amounts to a renormalization of the cohesive energy. When the value of $6B\kappa / k_B T$ decreases, the power-law tail broadens. When $6B\kappa / k_B T$ drops below one, c(S) cannot be normalized at the capsid formation threshold δμ = 0. In that case, we should expect to find a collection of aggregates with a very wide distribution in capsid sizes.

For the case of a *fluid* shell – i.e., the Helfrich Limit – but *with* non-zero spontaneous curvature, the equilibrium size distribution equals

$$c(S) \propto \exp^{\left\{4\sqrt{\pi}C_0 S^{1/2} - \left[\frac{1}{2}C_0^2 + \delta\mu/\kappa\right]S\right\}\kappa/k_BT} \tag{5.2}$$

using Eq. 3.4. This size distribution has a maximum at $S^* = 4\pi\left[\kappa C_0\right]^2 / \left[\delta\mu + \frac{1}{2}\kappa C_0^2\right]^2$, and the relative width $\left\langle \delta S^2 \right\rangle / S^{*2}$ of its maximum equals $k_BT / 8\pi\kappa$ in that regime. The location of the maximum depends on the chemical potential but it equals the (expected) value of $16\pi / C_0^2$ for δμ small compared to $\kappa C_0^2$. The peak in the distribution "survives" the introduction of elastic stress if the peak position S* is located below the buckling threshold, since the effect of elastic strain in that regime only amounts to a renormalization of the chemical potential (the peak position coincides in that regime with the minimum of the elastic energy shown in Fig.6). Above the buckling threshold, the peak in c(S) only survives if the (positive) *curvature* of the $\left(S_B / S\right)^{6B\kappa/k_BT}$ power law at S = S* is small compared to the (negative) curvature of the peak distribution $\propto \exp^{-\frac{1}{2}(S-S^*)^2/\left\langle \delta S^2\right\rangle}$. This condition is obeyed when $\left\langle \delta S^2 \right\rangle / S^{*2}$ is small compared to $k_BT / 6B\kappa$. Since $\left\langle \delta S^2 \right\rangle / S^{*2} = k_BT / 8\pi\kappa$ is actually *of the same order* for $B \approx 1.3$, it follows that the peak is either suppressed or seriously broadened. A numerical plot (not shown) demonstrates that the self-assembly peak is in fact completely suppressed by the addition of the elastic stress term.

**C) <u>Shape Degeneracy along the Sphere-to-Spherocylinder Transition Line: Encounter with Experiment</u>**

The transition between sphere and spherocylinder was found to be weakly first-order or continuous for FvK numbers near and above the buckling threshold. The elastic energy function E(S) does have minima as a function of S. The energy barrier separating them is however only of the order of 0.1κ, as shown in Fig.14b; this is of the order of the thermal energy $k_B T$, according to the estimates of Section 5A, which is too low to produce well-defined peaks in the c(S) function. We thus should expect to encounter extensive shape diversity at the transition line between sphere and spherocylinder in a self-assembly study of viral capsids. Note, from Fig.14B, that cones with m/n values close to one, such as m/n = 52/49, also are within a few $k_B T$ of the energy of the sphere and the spherocylinder.

Self-assembly studies of viral capsids that report both icosahedral and tubular structures as a function of physico-chemical control parameters – e.g. pH, salinity, or Ca++ ions – are available for CCMV[15] Alfalfa Mosaic Virus[17,37], the Polyoma/SV40 virus[38], and HIV-1[18,25]. The study of SV40 Virus reports[37] that pentamers efficiently assemble into shells in the presence of 1 M NaCl and 2 mM CaCl2 at neutral pH. At low temperatures and in the presence of ammonium sulfate, they form native T=7 capsids (see Fig.16A, with no genome molecules present). At room temperature and no ammonium sulfate, they form small T=1 icosahedral particles and tubular structures (Fig.16B and C). The tubular shells have a wide range of aspect ratios. Although this was not remarked upon by the authors, conical structures are in fact clearly visible as well, see Fig.16B. Finally, in the presence of 150 mM NaCl at pH 5, very long tubular shells appear.

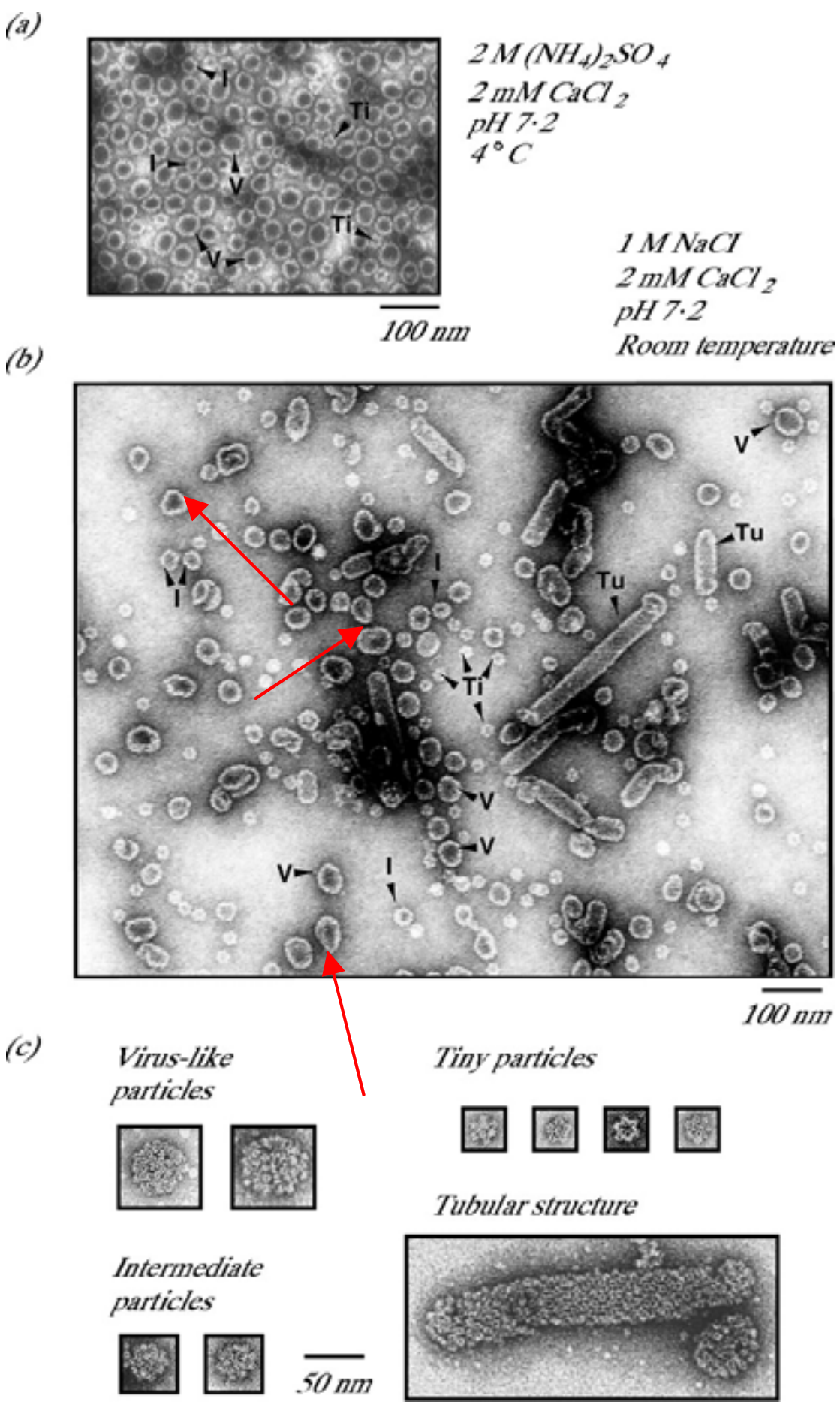

Fig.16: In vitro self-assembly of the VP1 capsid proteins of SV40 (from Ref. 38). Fig.16A: Electron micrograph of VP1 assembly in 2 M ammonium sulfate and 2 mM CaCl2 (pH 7·2) at 4 °C. V: Virus-sized shell; I, intermediate particle; Ti, tiny particle. Fig.16B: Electron micrograph of VP1 assembly in 1 M NaCl and 2 mM CaCl2 (pH 7·2) at room temperature. Tu, tubular structure. Arrow: conical structure. Fig.16C: Higher magnification of virus-sized shells, intermediate particles, tiny particles and a tubular structure are observed in (b).

Turning to CCMV, self-assembly studies of CCMV capsid proteins without genome molecules report that the native T=3 shell forms (20 nm diameter) for pH levels below 5.5 and for moderate salinity. At low ionic strength (near neutrality) and pH above 6, single and double-walled tubular shells form having diameters of 16 and 25 nm, respectively. In the transition region between the two structures, around 0.1 M salt and pH 5.0, shell shapes are indeed unstable with respect to ellipsoidal and, occasionally, conical shells. Finally, addition of viral RNA stabilizes the T=3 shell.

The structural degeneracy of SV40 and CCMV self-assembly appears to have no biological function, but this is not the case for AMV. An AMV self-assembly study reports that the shape of AMV capsid protein aggregates depends sensitively on the presence of single-stranded RNA. For instance, at a pH of 8.0 and in the presence of AMV RNA, spherical and ellipsoidal shells are observed of various length. No clear examples of conical AMV shells are seen. Self-assembly with more rigid (double-stranded) DNA molecules produces extremely long cylinders. This suggests that, *under natural conditions*, AMV is located near the transition line of the shape phase diagram of Fig.13. In the presence of RNA molecules, the interaction between the shell and the RNA would determine the actual morphology. This structural degeneracy is apparently exploited by the virus since the AMV genome consists of RNA molecules of various lengths that are packaged in different sized capsid shells. There thus appears to be no lack of evidence for a region of structural degeneracy in the self-assembly phase-diagrams of polymorphic viruses that is similar to the one encountered in the continuum theory. It should be noted however that the CCMV and AMV examples involve shells with a typical diameter of the order of 20 nm. It is a rather questionable assumption that the continuum theory can "work" in this regime and it would be interesting to investigate whether the structural degeneracy feature of the continuum theory will "survive" in a discrete description of small capsids.

In contrast, an example where continuum theory really is expected to be applicable are core particles of the HIV-1 virus. The immature HIV capsid is spherical, though not icosahedral, with diameters in the range of 120 to 260 nm[39]. After cleavage of the Gag capsid protein into CA ("capsid") and NC ("nucleocapsid") proteins – plus a matrix protein – the core reforms into a conical shell with a size of about 100 nm (majority case) plus a smaller fraction of tubular particles. The aperture angle of the cones – about 18 degrees – is consistent with a dominant 5-7 pentamer distribution. Solutions of viral RNA, CA and NC proteins readily self-assemble into conical and tubular shells very similar to the wild-type core particles. In the absence of viral RNA, conical shells form only at high salinity. Solutions of the CA protein by itself produce spherical and tubular shells, with a sharp transition taking place around pH 7.

This would indicate that CA protein shells are, under natural conditions, again located in the degeneracy region of Fig.13, with the actual shell structure being determined by RNA-protein interactions. Recent structural cryoTEM tomography studies[18] of individual HIV virions from a single infection show a dramatic polydisperisity in size and shape of nucleocapsids. Non-infectious virus-like-particles (VLPs) were produced in culture by introducing mutations in the reverse transcriptase and Rnase H enzymes and by preventing expression of the envelope protein. Tens of VLPs were selected for viewing along three orthogonal directions, as shown in Fig. 17, below:

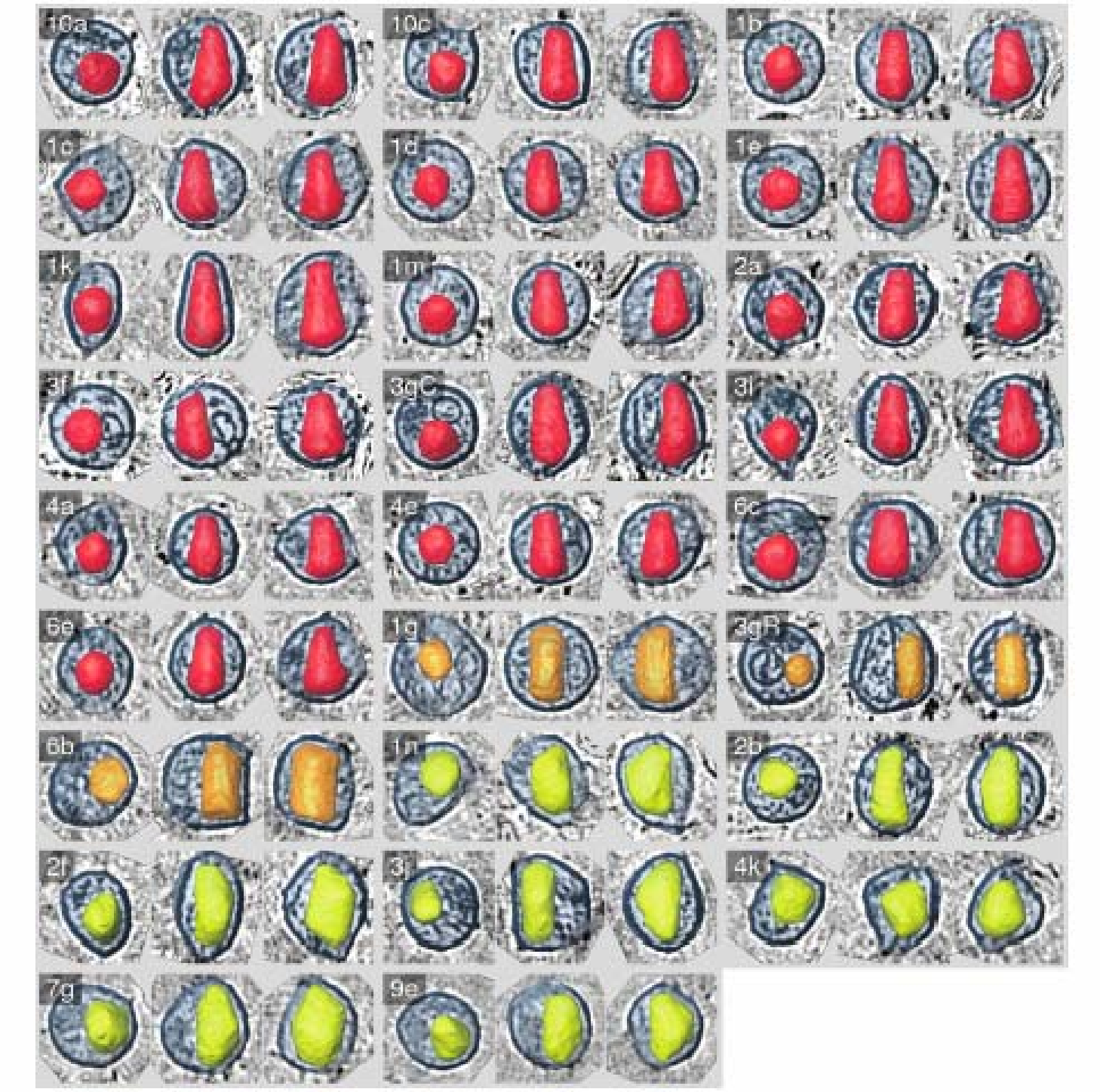

Fig. 17. CryoTEM tomography images of several tens of non-infectious HIV-1 virions, each shown along three orthogonal directions. A variety of conical, rodlike and other shapes are found from a single cell culture.

Note that, in addition to cones, there are many tubelike and irregular/globular shapes.

The FvK number of the (mature) cone is of the order of 20,000, well above the buckling threshold. According to our results, conical capsids with FvK numbers in this range *do* constitute a well-defined local minimum of the elastic energy with m/n ratios between 1.5 and 2 and spontaneous curvatures about twice the critical value (see Fig.15). These cones have a lower energy than the sphere but still a higher energy than spherocylindrical shells. One possibility for explaining how cones appear in spite of this energy ordering is that the spherocylinder has a lower volume than the cone with equal area, so self-assembly in the presence of the RNA genome might preclude formation of the spherocylindrical shells.

**D) Spontaneous Curvature versus Scaffolding**

A second point of confrontation between continuum theory and experimental studies of viral assembly concerns size-selection of capsids. We found in Section VB that spontaneous curvature could produce a well-defined peak in the concentration profile c(S) *only* for capsid sizes below the buckling threshold. For capsid sizes above the buckling threshold, size-selection by spontaneous-curvature is "spoiled" by the negative curvature of E(S) (see Fig.2). A second condition is that the dimensionless parameter $6B\kappa / k_B T$ has to be large compared to one. If these conditions are not met, then a separate size-selection mechanism must be operative, such as scaffolding. The second condition is certainly satisfied since $6B\kappa / k_B T$ is of the order of 100, using the estimates of Section VA.

We first recall that evidence for size control by spontaneous-curvature is available mostly for the smaller T=3 RNA viruses, which indeed never require scaffold structures for self-assembly. Next, the Hepatitis B virus and the Nudaurelia capensis $\omega$ virus[40] are examples of T=4 viruses whose capsid proteins assemble into T=4 shells without scaffold or genome molecules, both *in vitro* and in expression systems. The case of T=7 is more complex. For example, the T=7 double-stranded DNA phages in general rely on scaffold proteins for assembly[41], but in a somewhat equivocal manner. For instance, self-assembly of P22 capsid proteins without scaffold proteins actually *does* produce wild-type T=7 procapsids but also smaller T=4 shells[42] while self-assembly *with* scaffold proteins produces only T=7 shells. The P22 case is somewhat analogous to the SV40/Polyoma DNA virus in which case capsid proteins in solutions assemble into T=7 and T=1 caspids, as well as non-icosahedral 24 capsomer particles. In the presence of a condensed DNA genome, only the T=7 virus is formed. On the other hand, for the T=16 Herpes Simplex virus I the essential role of the scaffold during assembly is well-documented, e.g., a self-assembly study[43] of Herpes capsid proteins with a variable concentration of scaffold proteins reported that T=16 wild-type 100 nm diameter procapsids formed at higher scaffold protein concentrations. Below a critical concentration, the capsids collapsed to smaller shells with a 78 nm diameter. It does not appear to be known whether another T=16 virus, cytomegalovirus, or the T=25 adenovirus require a scaffold for assembly. The current data thus suggest that T=7 represents the borderline at which capsid assembly might no longer rely on spontaneous curvature as a mechanism for size control.

In summary, we have presented a shape phase-diagram for capsid self-assembly based on continuum elasticity theory, with dimensionless spontaneous curvature and the ratio of stretching to bending energies as the relevant degrees of freedom. The two main predictions include: (1) the existence of a degeneracy region in the shape phase diagram – at intermediate values of both the spontaneous curvature and the ratio of stretching to bending energies – where we find the simultaneous presence of cones, tubes, and spheres; and (2) a limit (around T=7) to the capsid size at which a monodisperse distribution is possible without scaffolding proteins playing a role. These behaviors are shown to be consistent with presently available experimental data.

We emphasize the central role played by the bending modulus of viral capsids and conclude that further work needs to be done on calculating and measuring this fundamental property. On the one hand, an experiment determining the spring constant of a large capsid, such as that of Herpes Simplex, for which continuum theory should apply, would be very helpful. On the other hand, a semi-empirical computation of this spring constant by a numerical simulation of an atomistic model would be useful as well.

Finally, while we do find the cone to appear as a minimum of the elastic shell Hamiltonian for intermediate values of the spontaneous curvature and the ratio of stretching to bending energies, it still has a higher energy than a spherocylinder of the same area. To reconcile this result with the experimentally observed fact that isometric HIV-1 cones appear as a more prevalent species, we must consider physical considerations that have not been included in the present theory. In particular, it is quite likely that the cone shape is stabilized relative to tubes and spheres through the role played by the viral RNA, i.e., by the interaction between this anionic polymer and the cationic N-termini of the proteins comprising the shell of the mature virion. Consideration of the kinetics of formation of the capsid aggregates is also likely to provide insights into the surprising prevalence of the conical shapes.

---

## References

[1] For an introduction to the structure and function of viruses, see: *Pinciples of Virology: Molecular Biology, Pathogenesis, and Control*, S. J. Flint et al. (ASM Press Washington, 2000).
[2] T. Douglas and M. Young, Nature **393**, 152 (1998).
[3] T. O. Yeates and J. E. Padilla, Curr. Opin. Struct. Biol. **12**, 464 (2002).
[4] T. S. Baker, N. H. Olson and S. D. Fuller, Microbiol. Mol. Biol. Rev. **63**, 862 (1999).
[5] H. Faenkel-Conrat and R. C. Williams, Proc. Nat. Acad. Sci. (USA) **41**, 690 (1955). See also, P. J. Butler and A. Klug, Proc. Natl. Acad. Sci. (USA) **69**, 2950 (1972).
[6] D. L. D. Caspar and A. Klug, *Cold Spring Harbor Symp. Quant. Biol.* **27**, 1 (1962).
[7] We will use "isometric" in this strict mathematical sense. The word is used in the virus crystallography literature to indicate viral shells that have the symmetry properties of an CK construction.
[8] J. Lidmar, L. Mirny and D. R. Nelson, Phys. Rev. **E68**, 051910 (2003).
[9] This definition of $\gamma$ differs from that of LMN by a factor of $4\pi$.
[10] R. Lata, J. F. Conway, N. Cheng, R. L. Duda, R. W. Hendrix, W. Wikoff, J. E. Johnson, H. Tsuruta and A. C. Steven, Cell **100**, 253 (2000).
[11] J. F. Conway, W. R. Wikoff, N. Cheng, R. L. Duda, R. W. Hendrix, J. E. Johnson and A. C. Steven, Science **292**, 744 (2001).
[12] M. F. Moody, Virology **26**, 567 (1965).
[13] N. A. Kiselev and A. Klug, J. Mol. Biol. **40**, 155 (19XY).
[14] X. F. Dong, P. Natarajan, M. Tihova, J. E. Johnson and A Schneemann, J. Virol. **72**, 6024 (1998).
[15] J. B. Bancroft, Advan. Virus Res. **16**, 99 (1970); K. W. Butler and P. J. G. Butler, J. Mol. Biol. **88**, 327 (1974).
[16] D. M. Salunke, D. L. D. Caspar and R. L. Garcea, Biophys. J. **56**, 887 (1989).
[17] S. Cusack, G. T. Oostergetel, P. C. Krijgsman and J. E. Mellema, J. Mol. Biol. **171**, 139 (1983).
[18] J. Benjamin, B. K. Ganser-Cornillos, W. F. Tivol, W. I. Sundquist and G. J. Jensen, J. Mol. Biol. **346**, 577 (2005).
[19] J. A. Speir, S. Munshi, G. Wang, T. S. Baker and J. E. Johnson, Structure **3,** 63 (1995).
[20] M. G. Rossmann and J. E. Johnson, Annu. Rev. Biochem. **58**, 533 (1989); J. E. Johnson, Proc. Nat. Acad. Sci. (USA) **93**, 27 (1996).
[21] C. Abad-Zapatero, S. S. Abdel-Meguid, J. E. Johnson, A. G. W. Leslie, I. Rayment, M. G. Rossmann, D. Suck and T. Tsukihara, Nature **286**, 33 (1980); A. J. Olson, G. Bricogne and S. C. Harrison, J. Mol. Biol. **171**, 61 (1983).
[22] R. C. Liddington, Y. Yan, J. Moulai, R. Sahli, T. L. Benjamin and S. C. Harrison, Nature **354**, 278 (1991).
[23] N. Funasaki, Adv. Colloid Interface Sci., **43**, 87 (1993).
[24] B. K. Ganser-Pornillos, U. K. von Schwedler, K. M. Stray, C. Aiken and W. I. Sundquist, J. Virol. **78**, 2545 (2004).
[25] J. A. G. Briggs, T. Wilk, R. Welker, H.-G. Kräusslich and S. D. Fuller, EMBO J. **22**, 1707 (2003).
[26] B. K. Ganser, S. Li. V. Y. Klishko, J. T. Finch, and W. I. Sundquist, Science, **283**, 81 (1999)

[27] *Micelles, Membranes, Microemulsions and Monolayers*, W.M. Gelbart, A. Ben-Shaul and D. Roux (Eds.) Springer, New York (1994); *Handbook of Biological Physics, Vol. 1, Structure and Dynamics of Membranes*, R. Lipowsky and E. Sackmann (Eds.), Elsevier, Amsterdam (1995).
[28] O. Y. Zhong and W. Helfrich, Phys. Rev. **A39**, 5280 (1989).
[29] A. E. Lobkovsky, Phys. Rev. E53, 3750 (1996).
[30] S. Seung and D. R. Nelson, Phys. Rev. **A38**, 1005 (1988).
[31] R. Zandi, D. Reguera, R. F. Bruinsma, W. M. Gelbart and J. Rudnick, Proc. Nat. Acad. Sci. (USA) **101**, 15556 (2004).
[32] Reddy VS, Giesing HA, Morton RT, Kumar A, Post CB, Brooks CL 3rd, Johnson JE. *Biophys J.* **74**, 546 (1998).
[33] P. Ceres and A. Zlotnick, Biochemistry (2002) **41** (11525)
[34] I. L. Ivanovska, P. J. de Pablo, B. Ibarra, G. Sgalari, F. C. MacKintosh, J. L. Carrascosa, C. F. Schmidt and G. J. L. Wuite, Proc. Nat. Acad. Sci. (USA) **101**, 7600 (2004).
[35] J.-P. Michel, I. Ivanovska, G. J. L. Wuite, C. F. Schmidt, M. G. Gibbons, W. S. Klug and C. M. Knobler, manuscript in preparation.
[36] L. D. Landau and E. M. Lifshitz, *Theory of Elasticity* Pergamom Press, Oxford (1988).
[37] G. Lebeurier, H. Fraenkel-Conrat, M. Wurtz and L. Hirth, Virology **43**, 51 (1971).
[38] S. Kanesashi, K. Ishizu, M. Kawano, S. Han, S. Tomita, H. Watanabe, K. Kataoka and H. Handa, J. Gen. Virol. **84**, 1899 (2003).
[39] T. Wilk, I. Gross, B. E. Gowen, T. Rutten, F. de Haas, R. Welker, H.-G. Krausslich, P. Boulanger and S. D. Fuller, J. Virol. **75**, 759 (2001).
[40] M. Canady, H. Tsuruta, and J. E. Johnson, J. Mol. Biol. **311**, 803 (2001).
[41] S. Casjens and R. W. Hendrix, in *The Bacteriophages, Vol. 1*, R. Calendar (Ed.), Plenum Press, New York (1988).
[42] P. A. Thuman-Commike, B. Greene, J. A. Malinski, J. King and W. Chiu, Biophys J. **74**, 559 (1998).
[43] W. W. Newcomb, F. L. Homa, D. R. Thomsen and J. C. Brown, J. Struct. Biol. **133**, 23 (2001).